\documentclass[%
 reprint,
superscriptaddress,
nofootinbib,
 amsmath,amssymb,
 aps,
]{revtex4-2}

\usepackage{graphicx} 
\usepackage{csquotes}
\usepackage{hyperref}
\usepackage{physics}
\usepackage{mathtools}
\usepackage[T1]{fontenc}
\mathtoolsset{showonlyrefs}
\hypersetup{colorlinks=true,linkcolor=blue,citecolor=blue,filecolor=blue,urlcolor=blue}
\begin{document}

\title{Exponential Scaling Barriers for Variational Quantum Eigensolvers}

\author{Manuel Hagelueken}
\email{manuel.hagelueken@ipa.fraunhofer.de}
\affiliation{Fraunhofer Institute for Manufacturing Engineering and Automation IPA, Nobelstraße 12, D-70569 Stuttgart, Germany}
\affiliation{Institute of Industrial Manufacturing and Management IFF, University of Stuttgart, Allmandring 35, Stuttgart, D-70569, Germany}

\author{David A. Kreplin}
\email{david.kreplin@hs-heilbronn.de}
\affiliation{Heilbronn University of Applied Sciences, Max-Planck-Str. 39, D-74076 Heilbronn, Germany}

\author{Florian Wieland}
\affiliation{Fraunhofer Institute for Manufacturing Engineering and Automation IPA, Nobelstraße 12, D-70569 Stuttgart, Germany}
\affiliation{Hochschule Esslingen (University of Applied Sciences), Kanalstraße 33, 73728 Esslingen am Neckar, Germany}

\author{Marco F. Huber}
\affiliation{Fraunhofer Institute for Manufacturing Engineering and Automation IPA, Nobelstraße 12, D-70569 Stuttgart, Germany}
\affiliation{Institute of Industrial Manufacturing and Management IFF, University of Stuttgart, Allmandring 35, Stuttgart, D-70569, Germany}

\author{Marco Roth}
\email{marco.roth@iao.fraunhofer.de}
\affiliation{Fraunhofer Institute for Manufacturing Engineering and Automation IPA, Nobelstraße 12, D-70569 Stuttgart, Germany}
\affiliation{Fraunhofer Institute for Industrial Engineering IAO, Nobelstraße 12, D-70569 Stuttgart, Germany}

\date{\today}
\begin{abstract} 
The Variational Quantum Eigensolver (VQE) is widely regarded as a promising algorithm for calculating ground states of quantum systems that are intractable for classical computers. This promise is typically motivated by the hope of mitigating the exponential growth of Hilbert space with system size. Here we scrutinize how the computational cost of adaptive VQE scales with the size of the target system. We demonstrate that the Rényi entropy derived from classical simulations predicts the required number of adaptive iterations of VQE with high accuracy (R$^2 \approx 0.99$). We validate this on a benchmarking set of more than 20 different molecules with active spaces ranging from four to ten orbitals. For these molecules, we find an exponential scaling of the number of adaptive iterations, and in turn, of the circuit depth with the system size. We therefore conclude that it is unlikely that VQE in its current form is able to simulate large molecular systems with high fidelity without exponential resource requirements.
\end{abstract}

\maketitle

\begin{figure*}[t]
\includegraphics[width=0.5\textwidth,clip]{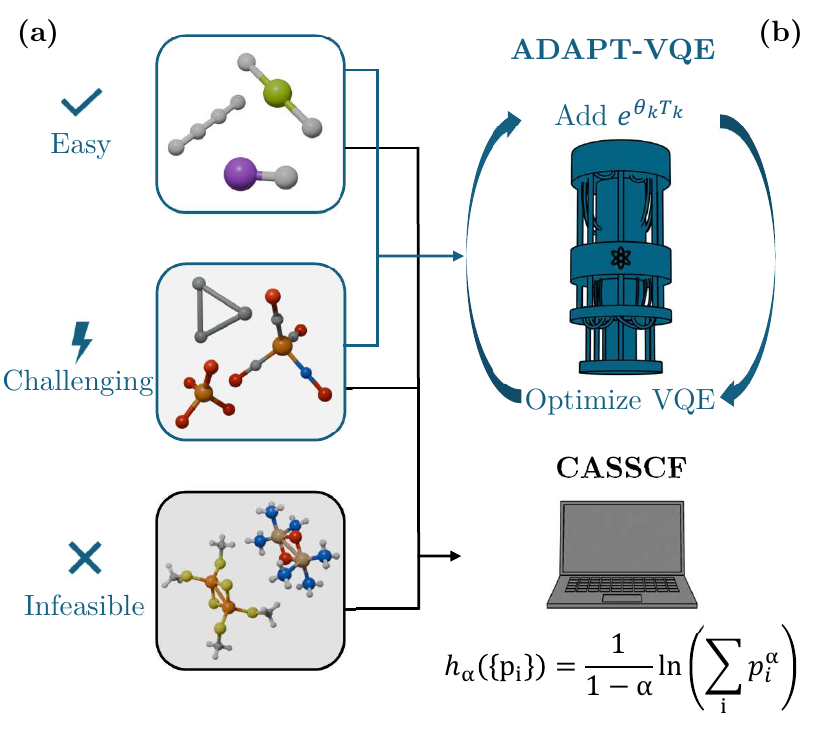}
\includegraphics[width=0.48\textwidth,clip]{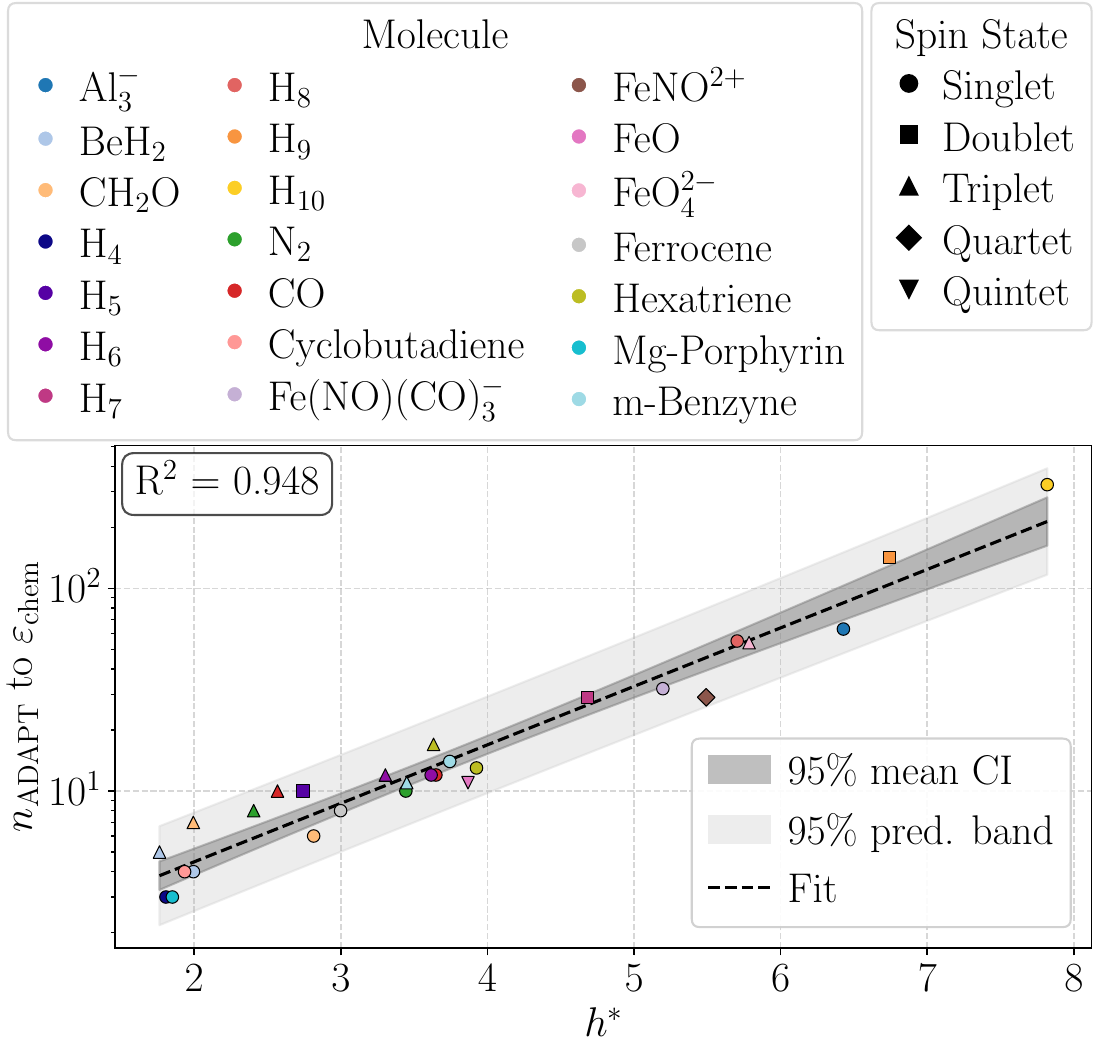}
\caption{\label{fig:paper_overview_sketch} (a) Schematic overview of the study. For each molecule in the benchmark set (Table~\ref{tab:molecule_set}), ADAPT-VQE simulations determine the number of iterations $n_\mathrm{ADAPT}$ required to reach a target energy error $\varepsilon$. The Rényi entropy $h_\alpha$, computed from the CI coefficient distribution of classical CASSCF calculations, serves as a complexity metric to predict $n_\mathrm{ADAPT}$. (b) Correlation between the Rényi entropy $h^\ast$ and $n_\mathrm{ADAPT}$ at chemical accuracy ($\varepsilon_\mathrm{chem}$) across the full benchmark set. Here, $h^\ast \equiv h_{\alpha^\ast}$ denotes the Rényi entropy at the order $\alpha^\ast$ that maximizes the coefficient of determination (R$^2$) at chemical accuracy. We find that $\alpha^\ast \approx 0.25$, yielding R$^2 = 0.948$ (see Section~\ref{sec:quantifying_problem_complexity}). The linear relationship in this semi-logarithmic plot implies that $n_\mathrm{ADAPT}$ scales exponentially 
with the Rényi entropy.}
\end{figure*}
\section{\label{sec:introduction}Introduction}
Accurate quantum-chemical simulations are key drivers of chemical innovation~\cite{Cao.2019}. Classical electronic-structure methods, such as density-functional theory~\cite{dft_ref} and coupled-cluster theory~\cite{coupled_cluster}, have been particularly successful in advancing molecular design. These methods usually employ approximations to keep the calculations computationally tractable. While approximations are sufficient for many molecules, they fail for several classes of systems, such as transition-metal clusters, which require accuracies beyond what approximate treatment can provide~\cite{iron_complex_classically_challenging,trans_metals_methods}. A common reason for this breakdown is strong static electron correlation, which cannot be described accurately by mean-field or single-reference methods. In such cases, the exact wave function is a superposition of many electronic configurations with comparable amplitudes, leading to pronounced multi-reference character. Treating such molecules requires resource-intensive multi-reference algorithms whose cost typically scales exponentially with the number of orbitals. Even with modern parallel implementations, exact multi-reference methods such as complete active space self-consistent field (CASSCF) and related approaches are usually limited to active spaces of about 18--20 electrons in 18--20 orbitals, and only a few larger cases have been demonstrated at very high computational cost~\cite{mcscf_limits}. Approximate multi-reference approaches such as the density-matrix renormalization group and full configuration-interaction quantum Monte Carlo extend this range but introduce uncontrolled errors or remain prohibitively expensive for many systems of chemical interest~\cite{dmrg_ref,fciqmc_ref,algo_quantum_chemistry_compare}.

Quantum-computing algorithms offer potential new pathways for treating these classically challenging systems~\cite{quantum_algos_promising_1,quantum_algos_promising_2,quantum_algos_promising_3}. The Variational Quantum Eigensolver (VQE) and its adaptive variants, collectively known as Adaptive Derivative-Assembled Pseudo-Trotter VQE (ADAPT-VQE), have emerged as promising near-term approaches~\cite{vqe_ref,adapt_vqe_ref}. These methods combine parametrized quantum circuits with classical optimization to approximate ground-state energies, requiring shallower circuits than fault-tolerant algorithms such as quantum phase estimation~\cite{qpe_ref}. Their modularity (encompassing ansatz design, operator pool construction, and optimizer choice) enables problem-specific tailoring but introduces trade-offs among expressivity, measurement cost, and convergence reliability~\cite{vqe_tradeoffs}.

While VQE and its variants have been applied to various molecular systems~\cite{ceo_vqe2025,scf_adapt_vqe,multiple_molecules_for_vqe}, a fundamental question remains open: Do these methods converge reliably for large, strongly correlated molecular systems and, if so, at what computational cost? We address this question by systematically analyzing the performance of ADAPT-VQE variants on a diverse benchmark set of 21 molecules spanning different basis sets, spin states, and active space sizes ranging from 4 to 10 orbitals (Table~\ref{tab:molecule_set}).

Our central finding is that the number of ADAPT iterations required to reach chemical accuracy scales exponentially with problem complexity. We demonstrate that the Rényi entropy, computed from purely classical CASSCF calculations (compare Figure~\ref{fig:paper_overview_sketch}), reliably predicts this computational cost, achieving coefficients of determination up to R$^2 = 0.99$. The linear relationship between Rényi entropy and the logarithm of required iterations implies exponential resource scaling. Using hydrogen chains as model systems, we confirm this exponential scaling directly (R$^2 = 0.98$) and show consistency between both prediction approaches. Extrapolating to larger systems, we estimate that chemically relevant molecules such as Cr$_2$, [2Fe-2S], and $[\mathrm{Cu}_2\mathrm{O}_2(\mathrm{NH}_3)_6]^{2+}$ require hundreds to thousands of ADAPT iterations to reach chemical accuracy, corresponding to circuit depths far beyond current hardware capabilities.

These findings have implications for VQE benchmarking. Commonly used test molecules such as H$_4$, LiH, and BeH$_2$~\cite{ceo_vqe2025,tetris_vqe_paper,qeb_pool_paper,qubit_pool_paper,adapt_vqe_ref,hardware_efficient_vqe,standard_benchmark_mols_limited_rep} exhibit low Rényi entropy and require few iterations, providing limited insight into performance on more complex systems. We therefore provide our curated benchmark set, MolVQE-21, as a community resource for more representative algorithm assessment.

The remainder of this paper is organized as follows. Section~\ref{sec:methods} presents the theoretical background and computational methodology, introducing the quantum algorithms, the Rényi entropy as a complexity metric, and the benchmark molecule set. Section~\ref{sec:results} presents our results: We quantify how iteration requirements vary across molecules (Section~\ref{sec:all_molecules}), establish the correlation between Rényi entropy and required iterations (Section~\ref{sec:quantifying_problem_complexity}), investigate scaling with system size (Section~\ref{sec:h_chains_equilibrium}), extrapolate to classically challenging systems (Section~\ref{sec:extrapolation}), and examine stretched geometries (Section~\ref{sec:stretched_h_chains}). We conclude with a discussion of the implications for the practical applicability of VQE-based methods.

\section{\label{sec:methods}Methods}
This section discusses the theoretical background of the concepts relevant in this work.

\subsection{\label{sec:vqe}Variational Quantum Eigensolver}

The Variational Quantum Eigensolver is a hybrid quantum-classical algorithm that approximates the ground-state energy of a given Hamiltonian $\hat{H}$ by minimizing the expectation value 
\begin{equation} 
\label{eq:energy_derivation}
E(\boldsymbol{\theta}) = \bra{\psi(\boldsymbol{\theta})} \hat{H} \ket{\psi(\boldsymbol{\theta})},
\end{equation}
where $\ket{\psi(\boldsymbol{\theta})}$ is a parametrized trial state prepared on the quantum computer. The variational principle guarantees that $E(\boldsymbol{\theta}) \geq E_0$ for all parameter choices, where $E_0$ is the true ground-state energy. Starting from an initial reference state $\ket{\psi_{\mathrm{init}}}$ (typically the Hartree-Fock determinant), the trial state is constructed by applying a parametrized unitary circuit, and the parameters $\boldsymbol{\theta}$ are iteratively updated using a classical optimizer to minimize $E(\boldsymbol{\theta})$ until $E(\boldsymbol{\theta})\approx E_0$.

A critical design choice in VQE is the structure of the parametrized ansatz circuit, which determines both the expressivity of the trial state and the circuit depth. The ansatz must be sufficiently flexible to represent the target ground state while remaining shallow enough for execution on near-term hardware. There are several ansatz templates used in the literature~\cite{hardware_efficient_vqe,qnp_vqe_ansatz,kupccgsd_vqe_ansatz}. Although most of our work is performed using ADAPT-VQE (see next section), we also include one specific ansatz to represent the behavior of fixed (non-adaptive) circuits: the perfect-pairing tiled unitary product states (pp-tUPS) ansatz~\cite{tUPS_paper}. This ansatz is motivated by generalized valence bond theory and connects bonding and anti-bonding orbital pairs through double excitations sandwiched by single excitations in a tiled structure. The circuit is organized in layers, where each layer applies one tile of paired excitations across the corresponding orbital pairs. Increasing the number of layers $L$ enhances expressivity at the cost of increased circuit depth and parameter count. This construction minimizes the number of required two-qubit gates per layer while maintaining chemical relevance, making it particularly well suited for describing bond-breaking processes. For the pp-tUPS ansatz, we reorder the spin-orbitals to pair bonding and anti-bonding orbitals as required by the ansatz structure.

While fixed-structure ansätze like pp-tUPS offer predictable circuit depths, they present optimization challenges that become more severe for deeper circuits or larger systems. Deep parametrized circuits are susceptible to barren plateaus, regions in parameter space where gradients vanish exponentially with system size, rendering optimization intractable~\cite{barren_plateaus_ref}. Additionally, the energy landscape of fixed ansätze often contains numerous local minima, causing optimizers to converge to suboptimal solutions that do not represent the true ground state~\cite{adapt_vqe_circumvents_vqe_opt_problems}. Although shallow ansätze like pp-tUPS mitigate the barren plateau problem, they remain susceptible to local minima.

Global optimization strategies can mitigate the local-minima problem. Basin Hopping Parallel Tempering VQE (BHPT-VQE)~\cite{tUPS_paper} combines multiple optimization trajectories at different effective temperatures, where higher temperatures increase the probability of accepting uphill moves to escape local minima. While BHPT-VQE can locate global minima for fixed-structure ansätze, it requires many independent VQE optimizations, and the computational cost generally grows prohibitively with system size due to the increasing number of local minima in the energy landscape.

\subsection{\label{sec:adapt-vqe}ADAPT-VQE}
To address the optimization challenges outlined in the previous section adaptive variants of VQE~\cite{adapt_vqe_ref} have been proposed. The idea is to construct the ansatz iteratively by appending operators selected from a predefined pool. This provides essentially a warm-start for the optimization in each iteration, avoiding random initialization, a well known source of barren plateaus \cite{adapt_vqe_circumvents_vqe_opt_problems}. Furthermore, in a noise-free environment, (ADAPT-)VQE is fully deterministic, which facilitates systematic benchmarking studies such as the present work.

The algorithm proceeds as follows. An operator pool $\{\hat{A}_j\}_{j=0}^{n_{\rm op}}$ of anti-Hermitian 
generators is defined, typically consisting of excitation operators motivated by 
coupled-cluster theory or hardware-efficient constructions. Each generator $\hat{A}_j$ 
induces a parameterized unitary operator
\begin{equation}
    \hat{U}_j(\theta) = e^{\theta \hat{A}_j},
\end{equation}
where $\theta \in \mathbb{R}$ is a free parameter. In the $n$-th ADAPT iteration, the energy gradient 
with respect to each operator is computed as
\begin{equation}
    g_{j,n} = \abs{\frac{\partial}{\partial \theta} E_{j,n}(\theta) \Big|_{\theta = 0}},
\end{equation}
where
\begin{equation}
    E_{j,n}(\theta) =
    \bra{\psi_{n}(\boldsymbol{\theta}_{n}^{\mathrm{opt}})}
      \hat{U}_j^\dagger(\theta)\, \hat{H}\, \hat{U}_j(\theta)
    \ket{\psi_{n}(\boldsymbol{\theta}_{n}^{\mathrm{opt}})}
\end{equation}
is the energy expectation value obtained by appending $\hat{U}_j(\theta)$ to the current 
optimized ansatz. Here, $\boldsymbol{\theta}_{n}^{\mathrm{opt}}$ denotes the optimized parameters from the preceding VQE step, and $\ket{\psi_0}$ is the initial reference state (typically the Hartree-Fock determinant). The operator $\hat{U}^\ast(\theta)$ with the largest gradient magnitude $g_{n}^\ast$ is appended to form the updated ansatz $\ket{\psi_{n+1}(\boldsymbol{\theta}_{n+1})}$, followed by a full VQE 
optimization of all parameters. This process repeats until the maximum gradient is below a predefined threshold or a predefined target error in the energy is reached. In the following, we denote the total number of gradient evaluation steps and subsequent VQE optimizations, i.e., the number of iterations in the outer ADAPT loop as $n_{\mathrm{ADAPT}}$.

The performance of ADAPT-VQE strongly depends on the choice of the operator pool. Different variants trade off the required $n_{\mathrm{ADAPT}}$, the total parameter count, and the circuit depth as measured by two-qubit (CNOT) gate count. Based on their success in previous studies~\cite{ceo_vqe2025}, we consider three variants in this work. Qubit-ADAPT-VQE~\cite{qubit_pool_paper} employs a hardware-efficient operator pool constructed by decomposing fermionic excitation operators into their constituent Pauli strings, each parametrized independently. This decomposition yields compact circuits with few CNOT gates per operator but typically requires more $n_{\mathrm{ADAPT}}$ and parameters. Qubit-excitation-based (QEB)-ADAPT-VQE~\cite{qeb_pool_paper} uses qubit excitation operators that preserve particle number without enforcing fermionic antisymmetry and spin symmetries, generally requiring fewer $n_{\mathrm{ADAPT}}$ and parameters at the cost of increased circuit depth per operator. CEO-ADAPT-VQE~\cite{ceo_vqe2025} introduces an operator pool consisting of operators, which combine multiple qubit excitation operators acting on the same set of qubits to further reduce the $n_{\mathrm{ADAPT}}$ and the CNOT gate count. To avoid losing expressivity in this approach an additional subroutine is added, which allows for multiple parameters in a single operator in the pool.

Standard ADAPT-VQE appends a single operator per iteration, which can lead to slow convergence for larger systems. The TETRIS extension~\cite{tetris_vqe_paper} addresses this limitation by appending multiple operators simultaneously, provided they act on disjoint qubit subsets. This parallelization significantly reduces $n_{\mathrm{ADAPT}}$ while maintaining comparable total parameter counts and circuit depths. Due to these favorable properties, we employ the TETRIS extension throughout this study unless stated otherwise.

For the VQE optimization within each ADAPT iteration, we employ the BFGS optimizer as implemented in Scipy~\cite{scipy_ref} combined with Hessian recycling~\cite{hessian_recycling}. Each inner VQE optimization was run to convergence with a gradient tolerance of $10^{-6}$. In this approach, the inverse Hessian matrix from the previous optimization is expanded with identity blocks for newly added parameters and used to initialize the subsequent optimization. Since parameters are initialized with their optimized values from the previous iteration, this strategy preserves curvature information across ADAPT iterations, reducing both classical computation time and the number of quantum measurements required for gradient evaluation. All quantum simulations in this work are performed as exact statevector simulations using Qulacs~\cite{qulacs_ref}.

\begin{table*}[t]
\centering
\caption{\label{tab:molecule_set}Molecules considered in this study with their active spaces, geometries, and computational details. The active space is specified as (electrons, orbitals). When given in the form [atomic orbitals] $\rightarrow$ (e, o), the active space was derived using the AVAS procedure with the indicated atomic orbitals as input. Bond lengths refer to nearest-neighbor distances. The spin column lists the electronic states studied for each molecule. Entries separated by \enquote{/} indicate that both states are included in the benchmark set. For hydrogen chains, the spin state depends on the electron count (singlet for even $n$, doublet for odd $n$). Molecules below the double line were used only for resource extrapolation (Section~\ref{sec:extrapolation}) and no ADAPT-VQE simulations were performed for these systems.}
\begin{tabular}{l c c c c c}
\hline
Molecule & Active space (e, o) & Geometry & Spin & Basis & Charge \\
\hline
\multicolumn{6}{l}{\textit{MolVQE-21 benchmark set (ADAPT-VQE simulations performed)}} \\
\hline
H$_n$ ($n=4,\dots,10$)  & ($n$,$n$) & 1.0~\AA & singlet or doublet & STO-3G & 0 \\
H$_n$ ($n=4,\dots,9$)         & ($n$,$n$) & 3.0~\AA & singlet or doublet & STO-3G & 0 \\
Cyclobutadiene & [C 2p$_\mathrm{z}$] $\rightarrow$ (4,4) & taken from~\cite{cyclobutadiene_ref} & singlet & cc-pVDZ & 0 \\
Mg-porphyrin   & (4,4) & taken from~\cite{kreplinphdthesis} & singlet & def2-TZVP & 0 \\
H$_6$          & (6,6) & 1.0~\AA & singlet/triplet & STO-3G/6-31G(d)/cc-pVTZ & 0 \\
m-Benzyne      & [C 2p$_\mathrm{z}$] $\rightarrow$ (6,6) & see Appendix~\ref{sec:molecular_geometry_optimization} & singlet/triplet & STO-3G/6-31G(d)/cc-pVTZ & 0 \\
BeH$_2$        & (4,6) & 1.3~\AA, taken from~\cite{ceo_vqe2025} & singlet/triplet & STO-3G/6-31G(d)/cc-pVTZ & 0 \\
N$_2$          & (6,6)~\cite{disco_vqe} & 1.2~\AA & singlet/triplet & STO-3G/6-31G(d)/cc-pVTZ & 0 \\
CH$_2$O        & (6,6) & taken from~\cite{ch2o_ref} & singlet/triplet & STO-3G/6-31G(d)/cc-pVTZ & 0 \\
CO             & [C 2p, O 2p] $\rightarrow$ (6,6) & 1.2~\AA & singlet/triplet & STO-3G/6-31G(d)/cc-pVTZ & 0 \\
Hexatriene     & [C 2p$_\mathrm{z}$] $\rightarrow$ (6,6) & see Appendix~\ref{sec:molecular_geometry_optimization} & singlet/triplet & STO-3G/6-31G(d)/cc-pVTZ & 0\\
Ferrocene      & [Fe 3d] $\rightarrow$ (10,7) & taken from~\cite{avas_paper} & singlet & cc-pVTZ-DK & 0 \\
Fe(NO)(CO)$_3^{-}$ & [Fe 3d] $\rightarrow$ (10,8) & taken from~\cite{avas_paper} & singlet & def2-TZVP & $-1$ \\
FeO$_4^{2-}$   & [Fe 3d] $\rightarrow$ (8,8) & taken from~\cite{avas_paper} & triplet & def2-TZVP & $-2$ \\
FeNO$^{2+}$    & (11,9)~\cite{feno2+_ref1} & taken from~\cite{feno2+_ref1,feno2+_ref2} & quartet & cc-pVTZ & $+2$ \\
FeO            & [Fe 3d, O 2p] $\rightarrow$ (12,9) & taken from~\cite{feo_ref} & quintet & def2-TZVP & 0 \\
Al$_3^{-}$     & [Al 3p] $\rightarrow$ (8,9) & taken from~\cite{al3_ref} & singlet & def2-TZVP & $-1$ \\
\hline
\hline
\multicolumn{6}{l}{\textit{Extrapolation targets (no ADAPT-VQE simulations; see Section~\ref{sec:extrapolation})}} \\
\hline
H$_{15}$       & (15,15) & 1.0~\AA & doublet & STO-3G & 0 \\
Cr$_2$         & (12,12)~\cite{cr2_reference} & 1.6788~\AA~\cite{cr2_reference} & singlet & cc-pwCVQZ-DK & 0 \\
{[2Fe-2S]}     & (14,12)~\cite{iron_complex_classically_challenging} & taken from~\cite{iron_complex_classically_challenging} & singlet & def2-TZVP & $-2$ \\
$[\mathrm{Cu}_2\mathrm{O}_2(\mathrm{NH}_3)_6]^{2+}$     & (16,16)~\cite{cu_complex_ref} & taken from~\cite{Cramer.2006} & singlet & def2-TZVPP & $+2$ \\
\hline
\end{tabular}
\end{table*}
\subsection{\label{sec:classical_reference}Classical Reference Calculations}

To establish reference energies for benchmarking VQE, we employ CASSCF calculations using PySCF~\cite{pyscf_ref} within the Born–Oppenheimer and non-relativistic approximations. As a starting point, we perform restricted Hartree–Fock (RHF) calculations for closed-shell states, yielding a single Slater determinant obtained self-consistently in a mean-field approximation. The RHF orbitals serve as the initial guess for the subsequent CASSCF optimization.

For weakly correlated systems, the true ground state is dominated by a single determinant, with only small contributions from excited configurations. Such dynamic correlation can be treated efficiently by classical methods like coupled-cluster theory. In contrast, strongly correlated systems exhibit multiple configurations with comparable amplitudes in the wave function, giving rise to pronounced multi-reference character that necessitates methods like CASSCF.

CASSCF partitions the molecular orbitals into three sets: doubly occupied core orbitals, a set of active orbitals with variable occupation, and empty virtual orbitals. Within the active space, the method performs a full configuration interaction (FCI) expansion while simultaneously optimizing both the orbital shapes and the configuration interaction (CI) coefficients. This treatment captures static (strong) correlation within the active space exactly, making CASSCF well suited for systems with significant multi-reference character.

The computational cost of CASSCF scales exponentially with the size of the active space due to the FCI treatment. Selecting an appropriate active space is therefore critical and non-trivial. In this study, we employ the atomic valence active space (AVAS) procedure~\cite{avas_paper} to guide active space selection based on the RHF orbitals.

The CASSCF calculation provides both the reference energy and the Hamiltonian for subsequent VQE simulations. Specifically, we extract the second-quantized Hamiltonian restricted to the active space from the converged CASSCF calculation. This active-space Hamiltonian is then mapped to a qubit representation using the Jordan–Wigner transformation~\cite{jordan_wigner_ref}, yielding the Hamiltonian $\hat{H}$ for VQE [Equation~\eqref{eq:energy_derivation}]. Consequently, the CASSCF energy represents the exact ground-state energy within the chosen active space, and the VQE energy error $\varepsilon = E_\mathrm{VQE} - E_\mathrm{CASSCF}$ quantifies the deviation from this target. We define chemical accuracy as $\varepsilon_{\mathrm{chem}} = 1.6 \times 10^{-3}~E_h$ (approximately 1~kcal/mol), which serves as our primary reference point for comparing algorithm performance across different systems.

\subsection{\label{sec:complexity_metric}Complexity Metric}

Since VQE typically starts from a single-determinant reference state (the Hartree–Fock solution), the difficulty of reaching the true ground state depends on how much the target wave function deviates from this single-reference picture. Systems with pronounced multi-reference character require the variational ansatz to build up contributions from many configurations, suggesting that multi-reference character can serve as a proxy for VQE problem complexity.

The CI coefficients $\{C_i\}_{i=1}^{N_{\mathrm{det}}}$ from a CASSCF calculation, where $N_{\mathrm{det}}$ is the number of determinants in the active space, provide a natural foundation for quantifying this complexity. These coefficients define a probability distribution $\{p_i\}_{i=1}^{N_{\mathrm{det}}}$, where $p_i = |C_i|^2$ is the probability associated with the $i$-th electronic configuration in the active space. A useful complexity metric should capture both the number of configurations with significant probability and the concentration of this distribution. Moreover, the relative importance of these aspects depends on the target value for $\varepsilon$: for modest energy accuracy requirements, only the dominant configurations matter, whereas high-accuracy calculations require faithful representation of the distribution's tail.

The Rényi entropy provides a unified framework that meets these requirements and is well established in quantum many-body physics~\cite{zhao_renyi_qmc,grover_renyi_fermions,renyi_entanglement_vqa,stab_renyi_magic}. For a probability distribution $\{p_i\}$ and a real parameter $\alpha > 0$, $\alpha \neq 1$, the Rényi entropy is defined as
\begin{equation}
    h_\alpha(\{p_i\}) = \frac{1}{1-\alpha} \ln\!\left( \sum_{i} p_i^\alpha \right).
    \label{eq:renyi_def}
\end{equation}
The parameter $\alpha$ controls sensitivity to different parts of the distribution: values $\alpha > 1$ emphasize large probabilities, while $0 < \alpha < 1$ increases sensitivity to smaller probabilities. In the limit $\alpha \to 0$, the Rényi entropy reduces to the logarithm of the number of non-zero probabilities. Additional special cases and their connections to established multi-reference diagnostics are discussed in Appendix~\ref{sec:renyi_special_cases}.

This tunability makes the Rényi entropy well suited for our analysis. When chemical accuracy is required, small values of $\alpha$ (below unity) appropriately weight contributions from the full distribution. For less stringent $\varepsilon$ requirements, larger $\alpha$ values focus the metric on the dominant configurations. 

We note that both the VQE optimization landscape and the Rényi entropy depend on the choice of orbital representation, since this choice influences the multi-reference character of a given molecular system. Crucially, the Rényi entropy reflects this basis dependence consistently: a basis that simplifies the CI expansion and therefore reduces the multi-reference character also simplifies the VQE optimization problem. Natural orbitals tend to minimize the multi-reference character, though this is not guaranteed in general~\cite{MR.diagnostics.2015}. Therefore in this study, we use natural orbitals throughout, which tend to provide favorable (lower) estimates of problem complexity. The generalization to other orbital representations is discussed in Appendix~\ref{sec:renyi_other_orb_reps}.

\begin{figure*}[t]
\includegraphics[width=0.99\textwidth,clip]{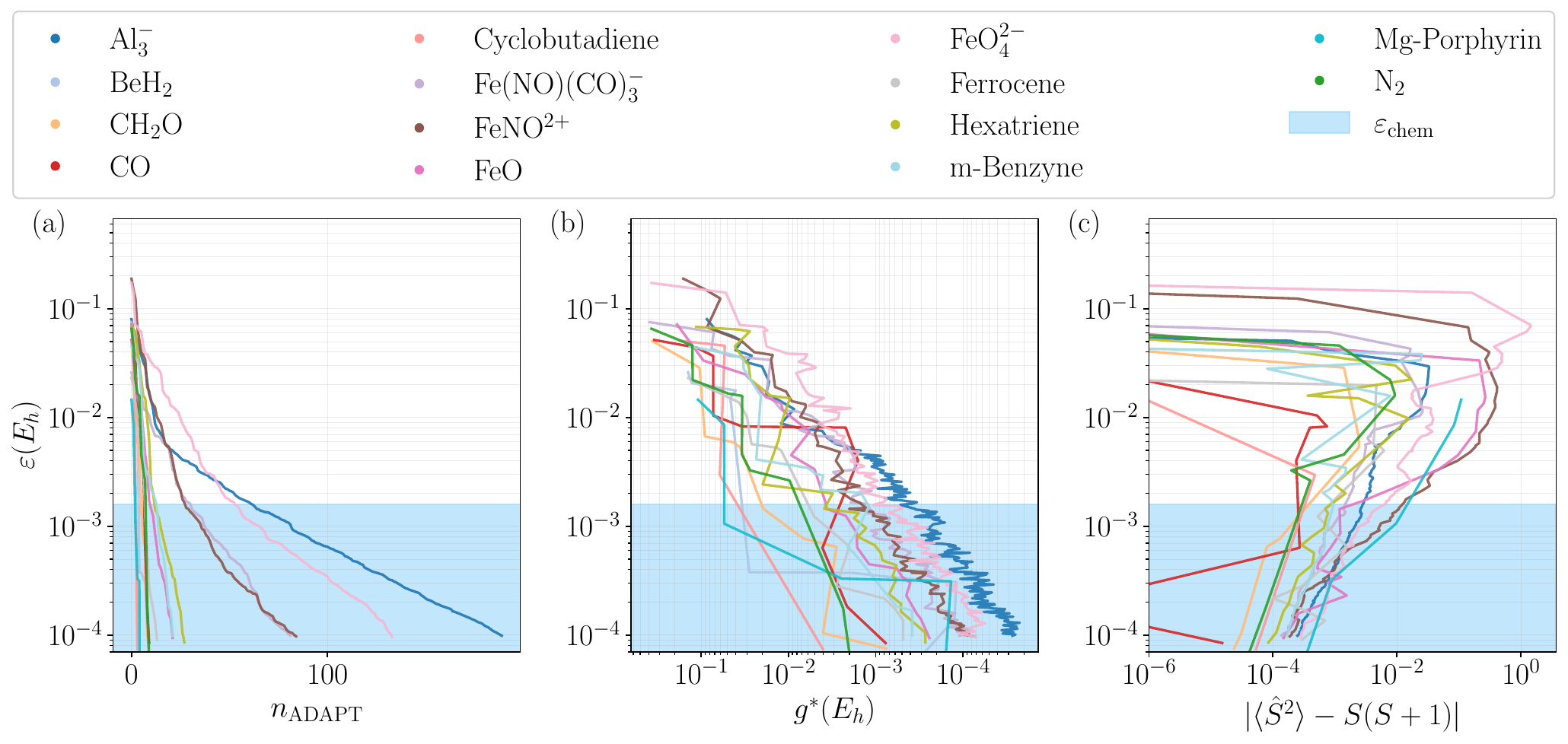}
\caption{\label{fig:all_other_molecules}
CEO-ADAPT-VQE with TETRIS extension applied to all molecules in Table~\ref{tab:molecule_set} except hydrogen chains. (a) Energy error $\varepsilon$ versus number of ADAPT iterations $n_{\mathrm{ADAPT}}$. (b) Energy error versus largest gradient magnitude $g^\ast$ among all operators in the pool at each iteration. (c) Energy error versus deviation $|\langle\hat{S}^2\rangle - S(S+1)|$ from the expected spin eigenvalue. The shaded region indicates $\varepsilon < \varepsilon_{\mathrm{chem}}$ (chemical accuracy).}
\end{figure*}
\subsection{\label{sec:molecule_set}Molecule Set}
To systematically assess VQE performance across different system types and sizes, we introduce MolVQE-21, a benchmark set of 21 molecular systems spanning active spaces from four to ten orbitals (Table~\ref{tab:molecule_set}). Complete molecular geometries and additional technical details are provided in the accompanying data repository~\cite{molvqe21_data}. The set is designed to cover a range of problem complexities and chemical diversity while remaining computationally tractable for ADAPT-VQE simulations.

The benchmark set comprises three categories. First, linear hydrogen chains (H$_4$--H$_{10}$) serve as model systems with tunable system size. Besides equilibrium geometries with interatomic distances of 1.0~\AA\ we consider hydrogen chains with stretched geometries (interatomic distance of 3.0~\AA) to investigate molecular systems exhibiting strong static correlation emerging from covalent bond breaking. In the literature covalent bond breaking systems are commonly considered as a proxy for classically challenging systems~\cite{ceo_vqe2025}. Second, small organic molecules (such as m-benzyne, hexatriene, and formaldehyde) provide chemically relevant test cases with six-orbital active spaces. For these systems, we consider multiple basis sets and spin states to assess the consistency of our findings. Third, transition-metal complexes (iron oxides, and related species) represent the most challenging category, exhibiting pronounced multi-reference character due to partially filled d-shells.

Except for the stretched hydrogen chains, all molecules are considered in their equilibrium configuration. Active space sizes were chosen to balance computational tractability with chemical accuracy, guided by AVAS recommendations and literature precedent where available. The resulting orbitals in the active spaces were checked for consistency. In our implementation, spin-orbitals are ordered by alternating alpha and beta electrons from lowest to highest energy.

\section{\label{sec:results}Results}
To evaluate the practical scaling limits of VQE for molecular systems, we systematically determine the resources required to reach chemical accuracy across the MolVQE-21 benchmark set. In Section~\ref{sec:all_molecules}, we quantify how $n_{\mathrm{ADAPT}}$ varies across molecules of different sizes and correlation strengths. In Section~\ref{sec:quantifying_problem_complexity}, 
we present our main result: the Rényi entropy computed from classical CASSCF calculations reliably predicts $n_{\mathrm{ADAPT}}$ for a given $\varepsilon$. In Section~\ref{sec:h_chains_equilibrium}, we establish the exponential scaling of $n_{\mathrm{ADAPT}}$ with system size using hydrogen chains as model systems and demonstrate consistency with the Rényi-based predictions. Section~\ref{sec:extrapolation} combines these findings to extrapolate $n_{\mathrm{ADAPT}}$ for classically challenging systems currently beyond direct simulation. Finally, Section~\ref{sec:stretched_h_chains} examines stretched hydrogen chains and explains why we focus in this study on equilibrium geometries for establishing scaling relationships.

\subsection{\label{sec:all_molecules}System Size Alone Does Not Predict Resource Requirements}
Based on systematic comparison of VQE variants (Appendix~\ref{sec:comparisons_vqe_variants}), 
we employ CEO-ADAPT-VQE with the TETRIS extension to all molecules in Table~\ref{tab:molecule_set}, treating hydrogen chains separately in Sections~\ref{sec:h_chains_equilibrium} and~\ref{sec:stretched_h_chains}, as 
CEO-ADAPT-VQE with the TETRIS extension requires the fewest iterations among tested methods. To keep simulation costs tractable for larger systems, we terminate each calculation upon reaching $\varepsilon \leq 10^{-4}~E_h$.

Figure~\ref{fig:all_other_molecules}(a) shows that $\varepsilon$ exhibits 
approximately linear convergence with $n_{\mathrm{ADAPT}}$ for all molecules. Several systems display two-phase behavior, with faster initial decay transitioning to a slower asymptotic rate. Crucially, the required $n_{\mathrm{ADAPT}}$ to reach a given $\varepsilon$ varies by more than an order of magnitude across molecules. While larger active spaces generally require more iterations, notable exceptions exist: For instance, Ferrocene (seven orbitals) reaches chemical accuracy in approximately 7 iterations, whereas Hexatriene (six orbitals) requires roughly 12 iterations, despite the smaller active space. This indicates that active space size, and consequently the number of qubits used in VQE alone is an insufficient predictor of ADAPT-VQE convergence, motivating the need for a complexity metric as discussed in Section~\ref{sec:quantifying_problem_complexity}.

We observe that the gradient magnitude $g^\ast$ [Figure~\ref{fig:all_other_molecules}(b)] decreases approximately linearly with $\varepsilon$ as $n_\mathrm{ADAPT}$ increases, indicating smooth convergence. Notably, for a given value of $\varepsilon$, $g^\ast$ is smaller for larger systems. The same trend is observed for hydrogen chains at equilibrium geometry (see Appendix~\ref{sec:h_additional_plots}). This can be explained by the fact that for larger systems deeper circuits (more $n_\mathrm{ADAPT}$) are required to reach the same $\varepsilon$. These deeper circuits apparently diminish the gradients.

To probe the properties of the state during optimization, we show the spin squared deviation $|\langle\hat{S}^2\rangle - S(S+1)|$ at each optimization step [Figure~\ref{fig:all_other_molecules}(c)]. In the beginning, the expected value $\langle \hat{S}^2 \rangle$  deviates significantly from the target value (by up to $\Delta \hat{S}^2 = \abs{\ev{\hat{S}^2}-S(S+1))} \approx 1$), indicating that early iterations select operators that temporarily break spin symmetry. In all cases, $\langle \hat{S}^2 \rangle$ converges back to within $10^{-3}$ of the expected value as the calculation proceeds. This transient symmetry breaking appears to be functionally important rather than incidental: In Appendix~\ref{sec:operator-pool_size}, we show that spin-preserving operator pools fail to converge for certain molecules where spin-non-preserving pools succeed. This indicates that temporary spin-symmetry breaking is required to navigate the optimization landscape toward the ground state.

\begin{figure}[t]
\includegraphics[width=0.49\textwidth,clip]{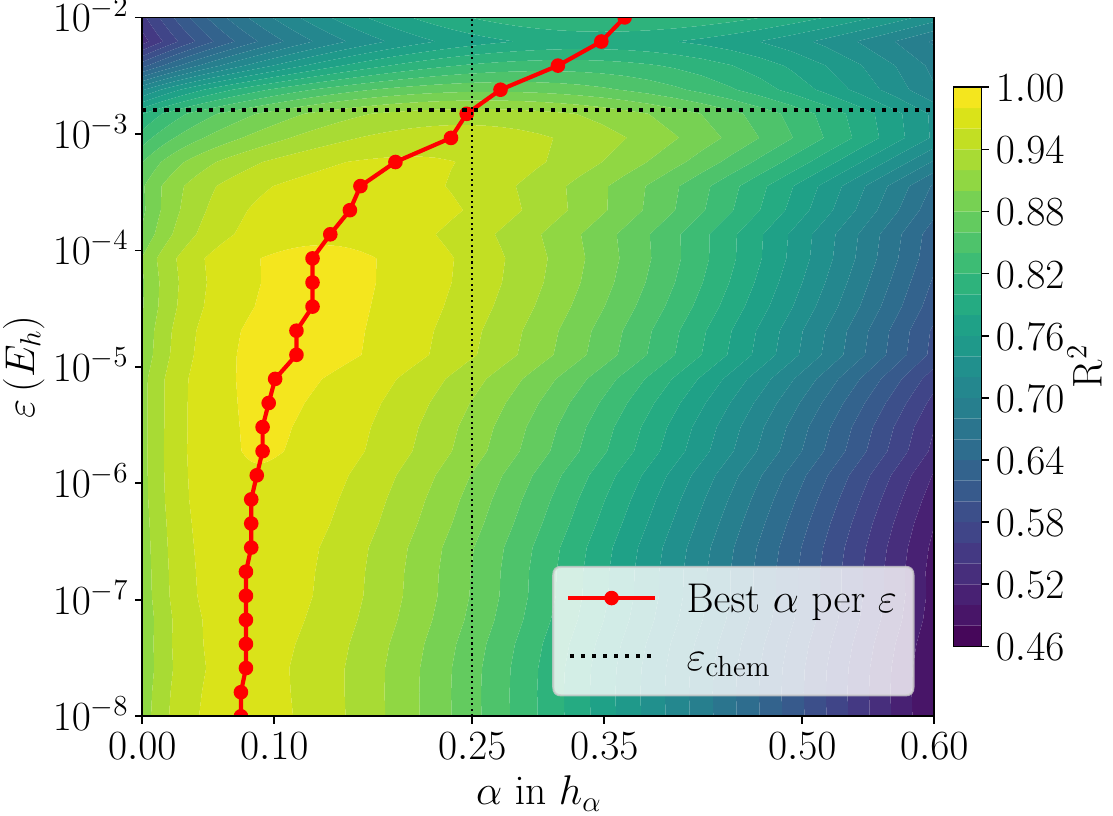}
\caption{\label{fig:r2_surface_renyi} Relationship between the Rényi entropy $h_\alpha$ and the required number of ADAPT iterations $n_{\mathrm{ADAPT}}$ to reach an energy error relative to the target state $\varepsilon$. The coefficient of determination (R$^2$) score of a linear fit between log($n_{\mathrm{ADAPT}}(\varepsilon)$) and $h_\alpha$ is shown for the seven six-orbital molecules described in Table~\ref{tab:molecule_set} in the given basis and spin states using QEB-ADAPT-VQE without TETRIS. The $\alpha$ resulting in the highest R$^2$ scores is indicated in red for each $\varepsilon$ and the best $\alpha$ at chemical accuracy $\varepsilon_{\mathrm{chem}}$ is marked (black).}
\end{figure}
\begin{figure*}[t]
\includegraphics[width=0.99\textwidth,clip]{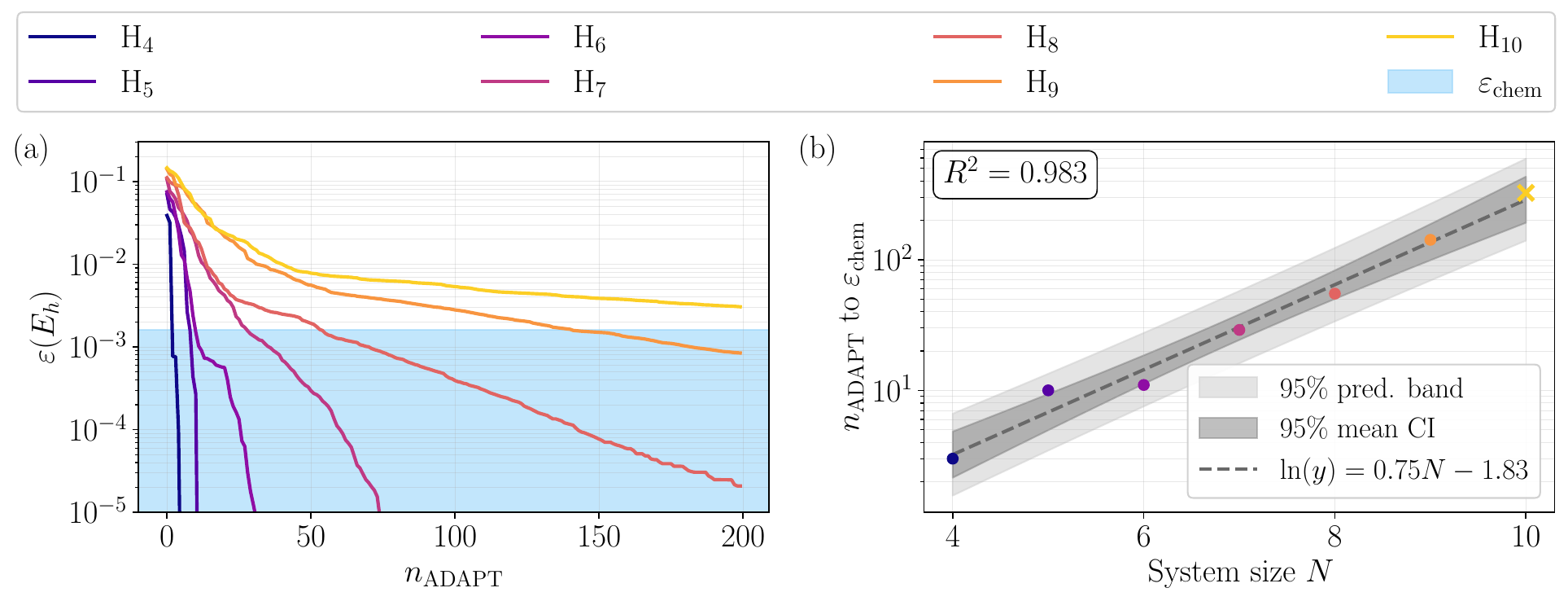}
\caption{\label{fig:Hn_chain_equi} Comparison of (a) the number of ADAPT iterations $n_{\mathrm{ADAPT}}$ and (b) the scaling of $n_{\mathrm{ADAPT}}$ required to reach chemical accuracy $\varepsilon_{\mathrm{chem}}$ as a function of system size $N$, for hydrogen chains consisting of four to ten atoms in equilibrium geometry (1.0~\AA\ interatomic distance). In panel (a), the vertical axis shows the energy error relative to the target state $\varepsilon$, with the chemical accuracy $\varepsilon_{\mathrm{chem}}$ highlighted. Panel (b) shows an exponential scaling behavior (black line) with the fit equation and coefficient of determination $R^2$ indicated. The data point for H$_{10}$ is marked with an \enquote{X} as it was obtained by extrapolation.}
\end{figure*}
\subsection{\label{sec:quantifying_problem_complexity} Quantifying Problem Complexity for VQE}
We now demonstrate that the Rényi entropy [Equation~\eqref{eq:renyi_def}] derived from CASSCF calculation serves as a complexity measure that predicts the computational cost of performing ADAPT-VQE on a given molecule.

We first calibrate the relationship between $h_\alpha$ and $n_{\mathrm{ADAPT}}$ using molecules with six-orbital active spaces (all six-orbital molecules from Table~\ref{tab:molecule_set} across all basis sets and spin states). This restricted set serves two purposes: Fixing the active space size isolates the effect of multi-reference character from trivial system-size dependence, and the moderate size permits convergence to small error ($\varepsilon = 10^{-8}~E_h$) at tractable computational cost. We then test whether the calibrated relationship generalizes to molecules of different sizes and chemical classes. 

For the calibration calculations, we employ QEB-ADAPT-VQE without the TETRIS extension, appending a single operator per iteration to maximize resolution in $n_{\mathrm{ADAPT}}$. We quantify the predictive capacity via the coefficient of determination (R$^2$) from linear regression of $\log(n_{\mathrm{ADAPT}})$ against $h_\alpha$. Figure~\ref{fig:r2_surface_renyi} shows R$^2$ as a function of both $\alpha$ and $\varepsilon$. The R$^2$ values form a smooth surface with a well-defined maximum for each error threshold. Three features merit discussion.

First, even the simple limit $\alpha \to 0$ (counting non-zero configurations) yields reasonable correlations (R$^2 > 0.8$ for $\varepsilon < \varepsilon_{\mathrm{chem}}$), but finite $\alpha$ values consistently improve the predictive capacity (R$^2 > 0.9$).

Second, the optimal $\alpha$ decreases systematically with decreasing $\varepsilon$ (red markers in Figure~\ref{fig:r2_surface_renyi}). This trend has a physical interpretation: ADAPT-VQE constructs the target state by iteratively appending excitation operators, each rotating probability amplitude between configurations. Since the magnitude of each probability can largely be absorbed into the rotation angle of the corresponding excitation, the number of ADAPT iterations scales primarily with the number of configurations that must be represented. For modest error requirements, only configurations with substantial probability matter, favoring larger $\alpha$. For small error reconstruction, even small-probability configurations must be captured, favoring smaller $\alpha$ that weight the distribution tail more heavily [see Equation~\eqref{eq:renyi_def}].

Third, for $\varepsilon > 10^{-2}~E_h$, the R$^2$ scores degrade because the required $n_{\mathrm{ADAPT}}$ values for different molecules all approach small values (often $n_{\mathrm{ADAPT}} < 5$), reducing the dynamic range for meaningful correlation. The highest R$^2$ values (up to $\approx 0.99$) are achieved towards lower $\varepsilon$, with the maximum occurring around $\varepsilon = 10^{-5}~E_h$.

At $\varepsilon_{\mathrm{chem}}$, the optimal value is $\alpha^\ast \approx 0.25$, yielding R$^2 = 0.94$. The R$^2$ maximum is broad, with $\alpha \in [0.2, 0.3]$ as visible from the contour structure in Figure~\ref{fig:r2_surface_renyi}. This suggests the relationship between the Rényi entropy and ADAPT-VQE's computational complexity is robust to the choice of $\alpha$, obviating the need for fine-tuning. In the remainder of this work, we denote the Rényi entropy at this optimal order as $h^\ast \equiv h_{\alpha^\ast}$, where $\alpha^\ast \approx 0.25$ maximizes R$^2$ at chemical accuracy. The corresponding slice through the landscape of Figure~\ref{fig:r2_surface_renyi} is shown in Appendix~\ref{sec:renyi_6_orbs_fit}. For the six orbital systems considered here, we observe that $n_{\mathrm{ADAPT}}$ varies from $\sim 8$ to $\sim 80$ iterations, underscoring the inadequacy of active space size alone as a complexity predictor.

Following the calibration on six-orbital molecules, we extend the analysis to the full MolVQE-21 
benchmark set (Table~\ref{tab:molecule_set}). Figure~\ref{fig:paper_overview_sketch}(b) shows $n_{\mathrm{ADAPT}}(\varepsilon_{\mathrm{chem}})$ versus the Rényi entropy $h^\ast = h_{0.25}$ for all molecules. To avoid over-representation, we include only the largest basis set (cc-pVTZ) for six-orbital molecules. Despite spanning active spaces from four to ten orbitals and diverse chemical classes, with $n_{\mathrm{ADAPT}}$ varying over two orders of magnitude, the data follow a consistent linear trend (R$^2 = 0.95$). Excluding the six-orbital calibration set yields R$^2 = 0.97$, confirming genuine predictive power. 

The linear relationship between $h_\alpha$ and $\log(n_{\mathrm{ADAPT}})$ has a critical implication: $n_{\mathrm{ADAPT}}$ scales exponentially with the Rényi entropy. Since circuit depth and parameter count scale approximately linearly with $n_{\mathrm{ADAPT}}$, this implies exponential growth in quantum resources with problem complexity as measured by $h_\alpha$.

The results in Figure~\ref{fig:paper_overview_sketch}(b) are obtained using CEO-ADAPT-VQE with the TETRIS extension, enabling tractable simulation of larger systems. In contrast to Figure~\ref{fig:r2_surface_renyi}, where a single operator is appended per iteration, here $n_{\mathrm{ADAPT}}$ is not equal to the number of operators appended (due to TETRIS parallelization). The consistency between results obtained with QEB-ADAPT-VQE (Figure~\ref{fig:r2_surface_renyi}) and CEO-ADAPT-VQE (Figure~\ref{fig:paper_overview_sketch}) with the TETRIS extension suggests that the Rényi entropy captures intrinsic problem complexity rather than method-specific behavior. A systematic $\alpha$ scan across all molecules (Appendix~\ref{sec:alpha_scan_all_molecules}) confirms that $\alpha \approx 0.25$ yields near-optimal R$^2$ for the full benchmark set independent of the operator pool and ADAPT-VQE variant.

The high R$^2$ scores (up to $\approx 0.99$ at $\varepsilon = 10^{-5}~E_h$; see Figure~\ref{fig:r2_surface_renyi}) establish the Rényi entropy as a robust predictor of $n_{\mathrm{ADAPT}}$. This is notable because $h_\alpha$, derived from the classical CI expansion, successfully predicts VQE performance despite the fundamentally different computational frameworks and the complex optimization landscape governing VQE convergence.

The optimal $\alpha$ depends on $\varepsilon$ but appears independent of the specific molecular system, basis set, and spin state. This universality suggests that the calibrated $\alpha$ values (Figure~\ref{fig:r2_surface_renyi}) can be applied to molecular systems beyond those studied here, enabling prediction of $n_{\mathrm{ADAPT}}$ for systems too large for direct ADAPT-VQE simulation, as demonstrated in Section~\ref{sec:extrapolation}.

\subsection{\label{sec:h_chains_equilibrium}Scaling with System Size: Hydrogen Chains}
The correlation between Rényi entropy and $n_{\mathrm{ADAPT}}$ established in Section~\ref{sec:quantifying_problem_complexity} enables predicting iteration counts across molecules of varying multi-reference character. However, it does not directly reveal how $n_{\mathrm{ADAPT}}$ scales with system size for a fixed class of molecules. To address this question, we analyze hydrogen chains H$_n$ ($n = 4$--$10$) at equilibrium geometry (1.0~\AA\ interatomic distance), which provide a homologous series with systematically increasing active space size while maintaining consistent electronic structure. We terminate calculations upon reaching $\varepsilon \leq 10^{-5}~E_h$ or $n_{\mathrm{ADAPT}} = 200$. We use CEO-ADAPT-VQE with the TETRIS extension as in previous sections.

Figure~\ref{fig:Hn_chain_equi}(a) shows $\varepsilon$ versus $n_{\mathrm{ADAPT}}$ for hydrogen chains of increasing length. All systems exhibit linear convergence as observed for the other equilibrium-geometry molecules (Section~\ref{sec:all_molecules}). This predictable convergence justifies extrapolation for H$_{10}$, where convergence to $\varepsilon_{\mathrm{chem}}$ was not achieved within our $n_{\mathrm{ADAPT}}$ limit(see Appendix~\ref{sec:extrapolation_h_10} for details).

\begin{figure}[t]
\includegraphics[width=0.49\textwidth,clip]{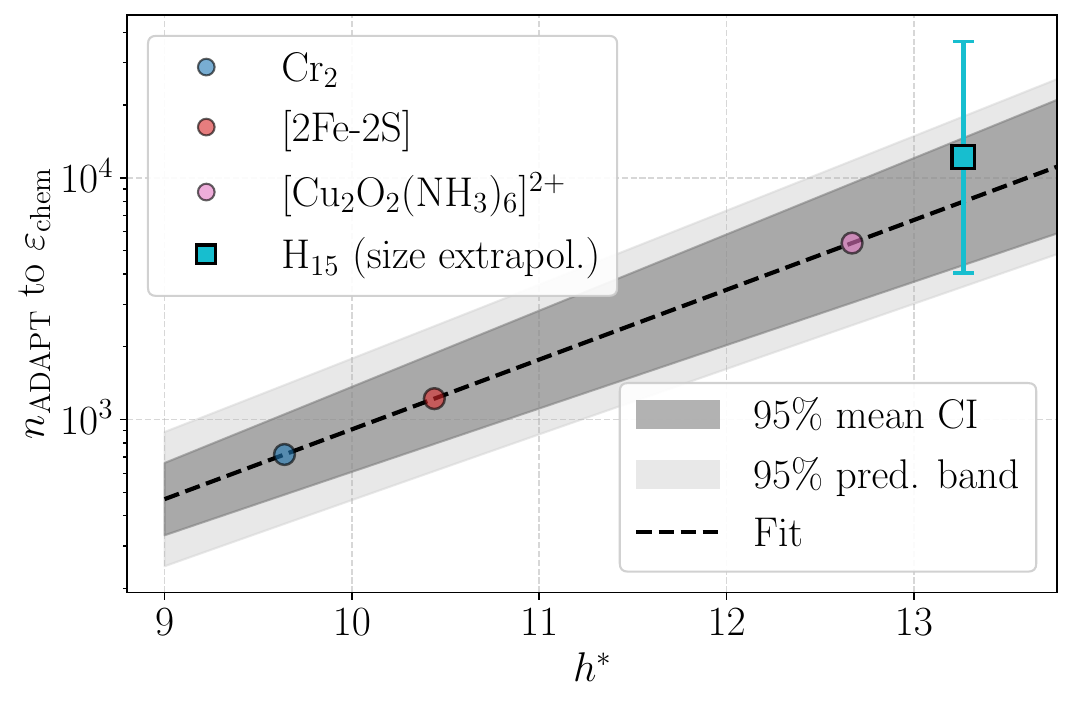}
\caption{\label{fig:renyi_extrapolation_only} Extrapolated number of ADAPT iterations $n_\mathrm{ADAPT}$ required to reach chemical accuracy ($\varepsilon_\mathrm{chem}$), estimated from the Rényi entropy $h^\ast$ using the linear fit in Figure~\ref{fig:paper_overview_sketch}(b). Here, $h^\ast \equiv h_{\alpha^\ast}$ denotes the Rényi entropy at the order $\alpha^\ast \approx 0.25$ that maximizes the coefficient of determination (R$^2$) at chemical accuracy. The \enquote{pred. band} indicates the 95\% confidence interval. For H$_{15}$, the square marker shows the independent estimate obtained from the system-size scaling fit in Figure~\ref{fig:Hn_chain_equi}(b).}
\end{figure}
Figure~\ref{fig:Hn_chain_equi}(b) presents a central result of the analysis: $n_{\mathrm{ADAPT}}(\varepsilon_{\mathrm{chem}})$ scales exponentially with the number of orbitals. A linear fit in the semi-log plot achieves excellent agreement (R$^2 = 0.98$), with the extrapolated H$_{10}$ point falling on the trend line. This exponential scaling is consistent with the theoretical analysis of Rényi entropy scaling (Appendix~\ref{sec:renyi_scaling}) and provides a second method to estimate $n_{\mathrm{ADAPT}}$ for larger hydrogen chains, which we compare against the Rényi-based prediction in Section~\ref{sec:extrapolation}. Since both CNOT gate count and parameter 
count scale approximately linearly with $n_{\mathrm{ADAPT}}$, circuit depth and total parameter count inherit this exponential scaling with system size.

\subsection{\label{sec:extrapolation}Extrapolation to Larger Systems}
The predictive relationship established in Section~\ref{sec:quantifying_problem_complexity} enables resource estimation for molecular systems beyond direct ADAPT-VQE simulation. To extrapolate $n_{\mathrm{ADAPT}}(\varepsilon_{\mathrm{chem}})$, we require only the Rényi entropy $h^\ast = h_{0.25}$, which can be computed from any classical method that provides the CI coefficients (here we use CASSCF).

We first validate our extrapolation approach by comparing two independent estimations for H$_{15}$: (i) system-size extrapolation from the H$_n$ scaling relationship [Figure~\ref{fig:Hn_chain_equi}(b)], and (ii) Rényi entropy extrapolation from the fit across all molecules [Figure~\ref{fig:paper_overview_sketch}(b)]. Figure~\ref{fig:renyi_extrapolation_only} shows, that both methods yield consistent estimates with overlapping 95\% confidence intervals, providing mutual validation of the extrapolation approach.

We then apply the Rényi-based extrapolation to molecules of practical interest: the chromium dimer Cr$_2$, the iron-sulfur cluster [2Fe-2S], and the binuclear copper complex $[\mathrm{Cu}_2\mathrm{O}_2(\mathrm{NH}_3)_6]^{2+}$ (see Table~\ref{tab:molecule_set} for details). These systems represent prototypical cases of strong multi-reference character and serve as common benchmarks for method development in quantum chemistry. While tractable with classical methods at the active space sizes considered here, they exemplify the correlation effects that become intractable in larger, chemically realistic models.

\begin{table}[t]
\centering
\caption{\label{tab:h15_extrapolation}Extrapolated resource estimates for H$_{15}$ to reach chemical accuracy ($\varepsilon_{\mathrm{chem}}$), comparing two independent extrapolation approaches: system-size scaling from H$_n$ chain data [Figure~\ref{fig:Hn_chain_equi}(b)] and Rényi entropy extrapolation [Figure~\ref{fig:renyi_extrapolation_only}]. The reported resources are the number of parameters $n_\mathrm{param}$ and CNOT gates $n_\mathrm{CNOT}$ in the final ansatz, estimated using per-iteration scaling factors from hydrogen chain simulations (Appendix~\ref{sec:cnot_parameter_estimation}). Values in brackets indicate 95\% confidence interval boundaries.}
\begin{tabular}{l c c}
\hline
Extrapolation method & $n_{\mathrm{param}} / 10^3$ & $n_{\mathrm{CNOT}} / 10^5$ \\
\hline
System-size     & 220 [95, 520]   & 9.3 [3.9, 22] \\
Rényi entropy   & 150 [64, 330]   & 6.1 [2.7, 14] \\
\hline
\end{tabular}
\end{table}
The extrapolated $n_{\mathrm{ADAPT}}(\varepsilon_{\mathrm{chem}})$ values for these systems are shown in Figure~\ref{fig:renyi_extrapolation_only}, ranging from approximately 700 for Cr$_2$ to over 5{,}000 for $[\mathrm{Cu}_2\mathrm{O}_2(\mathrm{NH}_3)_6]^{2+}$. Notably, these values are comparable to those of hydrogen chains with similar active space sizes, confirming that the exponential scaling relationship established for the model systems generalizes to chemically distinct molecules.

From the extrapolated $n_{\mathrm{ADAPT}}$ values and the average number of parameters and CNOT gates appended per iteration across smaller hydrogen chains (Appendix~\ref{sec:cnot_parameter_estimation}), we estimate the circuit resources required to reach chemical accuracy (Table~\ref{tab:h15_extrapolation}). Even the most optimistic estimates predict substantial resource requirements. For H$_{15}$, the lower confidence bound already requires approximately 64{,}000 parameters and 270{,}000 two-qubit gates.

These estimates have severe practical implications. For gradient-based optimization, each VQE iteration requires at least $2 \times n_{\mathrm{param}}$ circuit evaluations using the parameter-shift rule, multiplied by the number of shots $n_{\mathrm{shots}}$ needed to resolve shot noise and the number of measurement groups $n_{\mathrm{groups}}$ arising from non-commuting terms in the Hamiltonian. The circuit depth of $\mathcal{O}(10^5)$ CNOT gates imposes error-rate requirements far beyond current noisy intermediate-scale quantum hardware capabilities.

These extrapolations underscore the central finding of this work: the exponential scaling of $n_{\mathrm{ADAPT}}$ with problem complexity, as quantified by the Rényi entropy, presents a fundamental barrier to applying ADAPT-VQE to classically challenging molecular systems.

\begin{figure}[t]
\includegraphics[width=0.49\textwidth,clip]{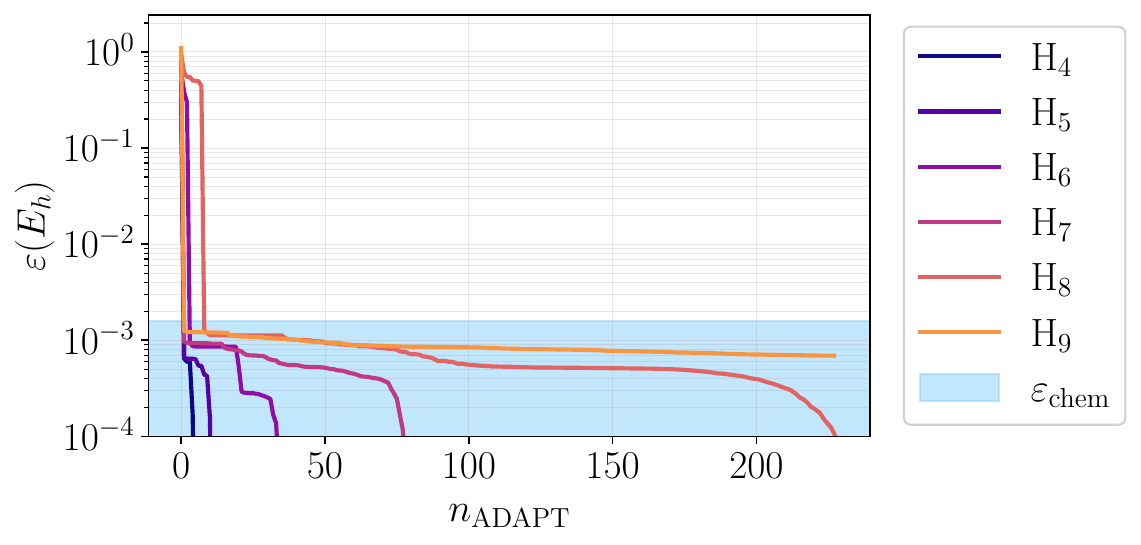}
\caption{\label{fig:Hn_chain_stretched_iter} Comparison of the number of ADAPT iterations $n_{\mathrm{ADAPT}}$ for hydrogen chains consisting of between four and nine atoms in stretched geometry (3.0~\AA\ interatomic distance). The y-axis shows the energy error relative to the target state $\varepsilon$, with the chemical accuracy $\varepsilon_{\mathrm{chem}}$ highlighted.}
\end{figure}
\subsection{\label{sec:stretched_h_chains}Stretched Hydrogen Chains and Equilibrium-Geometry Focus}
Stretched hydrogen chains are frequently used as model systems for strong static correlation, since bond elongation increases multi-reference character in a controlled manner~\cite{ceo_vqe2025}. Given our finding that the Rényi entropy predicts $n_{\mathrm{ADAPT}}$ for equilibrium geometries (Section~\ref{sec:quantifying_problem_complexity}), a natural question arises as to whether this relationship extends to stretched geometries. To address this, we examine ADAPT-VQE convergence for hydrogen chains at 3.0~\AA\ interatomic distance, comparing the behavior to the equilibrium results of Section~\ref{sec:h_chains_equilibrium}. We use CEO-ADAPT-VQE with the TETRIS extension as in previous sections.

Figure~\ref{fig:Hn_chain_stretched_iter} shows $\varepsilon$ versus $n_{\mathrm{ADAPT}}$ for stretched H$_n$ ($n = 4$--$9$), with simulations terminated at $\varepsilon \leq 10^{-4}~E_h$ or $n_{\mathrm{ADAPT}} = 230$. In contrast to the monotonic linear convergence observed at equilibrium geometries, stretched chains exhibit step-wise, non-monotonic convergence. Chemical accuracy is reached within the first few iterations for all system sizes, yet convergence to smaller error requires substantially more iterations than in the equilibrium case. Similar step-wise behavior has been reported in other studies of stretched geometries~\cite{ceo_vqe2025}.

This qualitative difference can be understood from the underlying physics. In stretched hydrogen chains, the system approaches the dissociation limit where a product state of nearly independent atoms lies energetically close to the true ground state. ADAPT-VQE rapidly identifies such a low-energy configuration (as evidenced by reaching $\varepsilon_{\mathrm{chem}}$ within the first few iterations) but this state has incorrect spin character: $\langle \hat{S}^2 \rangle$ deviates significantly from the expected eigenvalue even after chemical accuracy is reached (Appendix~\ref{sec:h_additional_plots}). The subsequent plateau reflects the difficulty of rotating from this energetically favorable but symmetry-broken state toward the true spin-eigenstate. In contrast, transition-metal complexes at equilibrium geometries exhibit multi-reference character arising from near-degenerate d-orbitals and metal-ligand interactions, where no such trivial low-energy state exists. The ground state must therefore be constructed systematically from the outset, resulting in the monotonic convergence observed in Section~\ref{sec:all_molecules}. This behavior underscores the value of monitoring $\langle \hat{S}^2 \rangle$ alongside energy during VQE optimization.

These findings have implications for VQE benchmarking. Commonly used test molecules such as H$_4$, BeH$_2$, and LiH at equilibrium appear in the lower-left region of Figure~\ref{fig:paper_overview_sketch}(b), requiring few ADAPT iterations due to their weak multi-reference character (LiH reaches $\varepsilon_{\mathrm{chem}}$ in a single iteration with a (2,4) active space). While stretched geometries produce strong multi-reference character ($h^\ast$ ranges from approximately 2.6 for H$_4$ to 8.4 for H$_9$), their step-wise convergence behavior differs qualitatively from that of the equilibrium transition-metal complexes in our benchmark set. For sufficiently stringent error thresholds ($\varepsilon \ll \varepsilon_{\mathrm{chem}}$), the Rényi entropy again becomes predictive of $n_{\mathrm{ADAPT}}$ (Appendix~\ref{sec:renyi_stretched}), but the intermediate-accuracy regime relevant for practical applications exhibits irregular behavior that complicates systematic analysis. Equilibrium geometries of chemically diverse molecules thus provide a more reliable basis for establishing scaling relationships and extrapolating to larger systems.

\section{\label{sec:discussion}Discussion}
In this work, we investigate the ability of VQE-based methods to reliably converge for large, strongly correlated molecular systems and determine the associated computational cost. To address this question, we assembled the MolVQE-21 benchmark set of 21 molecules spanning active spaces from four to ten orbitals with varying degrees of multi-reference character.

Our central finding is that the number of ADAPT iterations $n_{\mathrm{ADAPT}}$ required to reach chemical accuracy scales exponentially with system size. Since circuit depth and parameter count scale approximately linearly with $n_{\mathrm{ADAPT}}$, this implies exponential growth in quantum resources, which is a fundamental barrier for VQE-based approaches.

We established this scaling through complementary analyses. First, we showed that the Rényi entropy $h_\alpha$, computed from classical CASSCF calculations, predicts $n_{\mathrm{ADAPT}}$ across chemically diverse molecules with high fidelity (R$^2$ up to 0.99). The appropriate order of the Rényi is robust across a physically motivated range of values and depends on the target error but is independent of the molecular system, enabling resource estimation for molecules beyond direct quantum simulation. Second, using hydrogen chains as model systems with systematically increasing size, we demonstrated an exponential relationship between active space size and $n_{\mathrm{ADAPT}}(\varepsilon_{\mathrm{chem}})$ (R$^2 = 0.98$). The consistency between these two approaches provides mutual validation. Extrapolating to larger systems, we estimate that reaching chemical accuracy for H$_{15}$ requires at minimum $3{,}600$ ADAPT iterations, corresponding to $64{,}000$ parameters and $270{,}000$ CNOT gates. For chemically relevant molecules of similar size such as $[\mathrm{Cu}_2\mathrm{O}_2(\mathrm{NH}_3)_6]^{2+}$, the estimated ADAPT iteration count is of similar order of magnitude.

Our analysis also informs benchmarking practices. Commonly used test molecules (H$_4$, LiH, BeH$_2$) exhibit low Rényi entropy and require few ADAPT iterations, while stretched geometries, though strongly correlated, show qualitatively different convergence behavior than equilibrium transition-metal complexes. To address this mismatch, we introduce MolVQE-21 as a benchmark suite explicitly designed to enable more realistic and representative assessment of quantum algorithms.

Several known challenges of VQE were not the focus of this study but merit brief discussion. The measurement overhead in VQE is substantial: Shot noise requires repeated measurements scaling quadratically with the inverse precision, the number of Pauli strings in molecular Hamiltonians scales as $\mathcal{O}(N^4)$, and ADAPT-VQE additionally requires gradient evaluation over the operator pool in each iteration. Various strategies address these issues, including measurement grouping~\cite{classical_shadow_derandom}, operator pool reduction~\cite{mcp_paper}, classical shadows~\cite{classical_shadow}, and gradient reuse techniques~\cite{reuse_vqe_measurements,reuse_vqe_measurements_2}. Similarly, optimization efficiency can be improved through Hessian recycling~\cite{hessian_recycling}, circuit pruning~\cite{pruned_adapt_vqe}, and overlap-based methods~\cite{overlap_adapt_vqe}.

While these VQE extensions are effective in their domain, they do not mitigate the fundamental scaling problem identified in this work. The TETRIS extension~\cite{tetris_vqe_paper} and improved operator pools such as CEO-ADAPT-VQE~\cite{ceo_vqe2025} reduce $n_{\mathrm{ADAPT}}$, but our results show that even these optimized variants exhibit exponential scaling with system size. Classical pretraining schemes~\cite{classical_pretraining,tensornetwork_pretraining} could potentially initialize the optimization closer to the target state, but for systems where classical pretraining is effective, it remains unclear whether subsequent quantum optimization provides additional value.

Even if these additional challenges of VQE were resolved, the exponential scaling of $n_{\mathrm{ADAPT}}$ leads to an exponential overhead in the optimization, which also renders quantum-inspired classical methods based on VQE impractical. We emphasize that this does not imply VQE is entirely without value. It may serve as a state preparation subroutine for more sophisticated algorithms such as quantum phase estimation or for extracting properties from approximate ground states~\cite{learn_on_gs_wf}. However, our findings challenge the claim that (ADAPT-)VQE in its current form alone can tackle classically intractable molecular systems to chemical accuracy and beyond.
\\
\section*{Data Availability}
The data required to reproduce the figures in this study are hosted at \href{https://doi.org/10.5281/zenodo.19051535}{https://doi.org/10.5281/zenodo.19051535}.

\section*{Code Availability}
The underlying code for this study is not publicly available due to export control restrictions but may be made available to qualified researchers upon reasonable request from the corresponding author.

\begin{acknowledgments}
This work was supported by the German Federal Ministry of Research, Technology and Space through the project H2Giga-DegradEL3 (grant no. 03HY110D). The authors disclose the use of LLM-based tools for grammar and spell-checking.
\end{acknowledgments}

\bibliography{bibliography}

\appendix

\begin{figure*}[t]
\includegraphics[width=0.99\textwidth,clip]{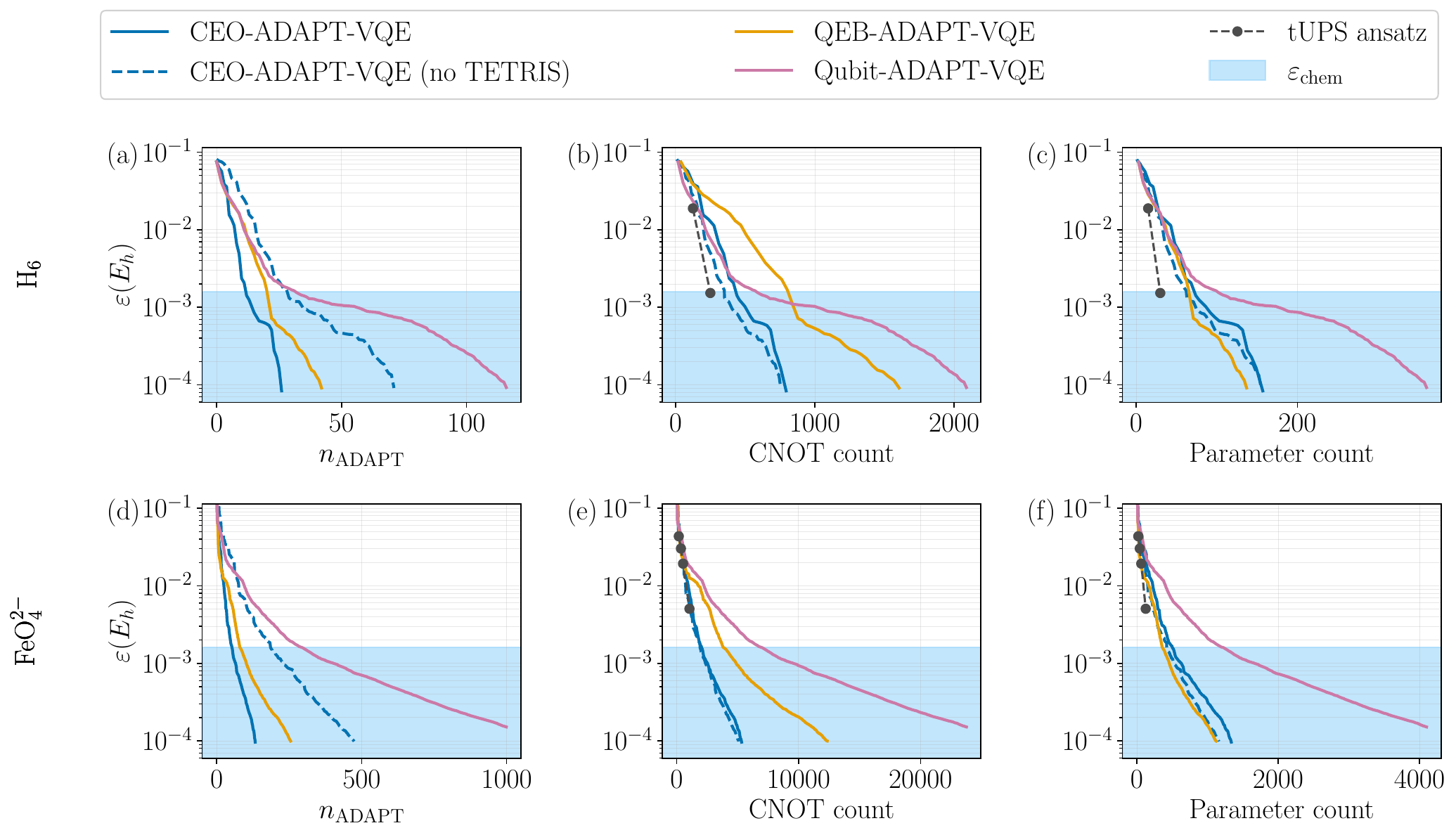}
\caption{\label{fig:vqe_variant_comparison} Comparison of VQE variants for H$_6$ (top row, 1.0~\AA, singlet) and FeO$_4^{2-}$ (bottom row, triplet). Columns show (left) number of ADAPT iterations $n_{\mathrm{ADAPT}}$, (middle) CNOT gate count, and (right) parameter count, all as functions of energy error $\varepsilon$. The shaded region indicates $\varepsilon < \varepsilon_{\mathrm{chem}}$ (chemical accuracy). For pp-tUPS, each point corresponds to a fixed number of layers $L$.}
\end{figure*}
\section{\label{sec:comparisons_vqe_variants}Comparison of VQE Variants}
The results presented in this study primarily employ CEO-ADAPT-VQE with the TETRIS extension, with the exception of the six-orbital calibration in Section~\ref{sec:quantifying_problem_complexity}, which uses QEB-ADAPT-VQE without TETRIS to maximize resolution in $n_{\mathrm{ADAPT}}$ (one operator appended per iteration). Here we justify this choice by systematically comparing VQE variants on two representative molecules: H$_6$ at equilibrium geometry (1.0~\AA), a common benchmark in the literature~\cite{ceo_vqe2025}, and FeO$_4^{2-}$, which exhibits the largest Rényi entropy among molecules with fewer than nine orbitals in our benchmark set [Figure~\ref{fig:paper_overview_sketch}(b)] and thus represents a more challenging test case.

Figure~\ref{fig:vqe_variant_comparison} compares Qubit-ADAPT-VQE, QEB-ADAPT-VQE, CEO-ADAPT-VQE, and BHPT-VQE with the pp-tUPS ansatz across three metrics: $n_{\mathrm{ADAPT}}$, CNOT gate count, and parameter count. For CEO-ADAPT-VQE, we show results both with and without the TETRIS extension, as this combination has not been as extensively studied as TETRIS with Qubit-ADAPT-VQE or QEB-ADAPT-VQE. The TETRIS extension increases CNOT and parameter counts slightly but reduces $n_{\mathrm{ADAPT}}$ substantially, motivating its use for calculations on larger systems.

Among the ADAPT variants, CEO-ADAPT-VQE requires the fewest iterations and CNOT gates, followed by QEB-ADAPT-VQE. This ordering reflects the design of the CEO operator pool, which combines multiple double excitations acting on the same qubits into single multi-parameter operations, reducing both iteration count and gate overhead while preserving expressivity~\cite{ceo_vqe2025}. The trade-off is a modest increase in parameter count and, consequently, more optimizer iterations per ADAPT step to reach convergence. While the gradient evaluation for all three operator pools scales as $\mathcal{O}(n_q^5)$~\cite{pool_scaling}, where $n_q$ is the number of qubits, the CEO pool contains fewer operators in absolute terms, partially offsetting the per-iteration measurement cost.

The pp-tUPS ansatz with BHPT-VQE optimization achieves lower CNOT and parameter counts than the ADAPT variants for a given $\varepsilon$. However, this comparison requires important caveats. First, each pp-tUPS data point represents a fixed number of layers $L$, and global optimization provides no guarantee of convergence to the true global minimum; these energies therefore represent upper bounds on what the ansatz can achieve. Second, the global optimization requires many independent VQE runs, incurring measurement overhead that can exceed practical limits~\cite{ceo_vqe2025}. Third, unlike ADAPT-VQE, which systematically builds expressivity, fixed-depth ansätze offer no guarantee that the target state lies within the ansatz manifold for a given $L$. 

These limitations manifest concretely for FeO$_4^{2-}$: despite substantial computational effort, we did not achieve chemical accuracy with BHPT-VQE. Extrapolating the observed trend in Figure~\ref{fig:vqe_variant_comparison} suggests that approximately $L = 9$ layers would be required.

These findings are consistent with previous benchmarks on small molecules~\cite{ceo_vqe2025} and, importantly, extend to the more challenging FeO$_4^{2-}$ system, supporting the use of CEO-ADAPT-VQE with the TETRIS extension for the larger molecules in our benchmark set. We consider this variant the most promising for larger systems: it minimizes $n_{\mathrm{ADAPT}}$ without requiring prohibitive global optimization. Consequently, the scaling limitations identified in this study represent a conservative estimate of the challenges facing VQE-based approaches; other variants would require equal or greater resources.

\section{\label{sec:renyi_special_cases}Special Cases of the Rényi Entropy}
The Rényi entropy interpolates between several well-known information-theoretic quantities depending on the order $\alpha$. Understanding these limiting cases clarifies why intermediate values of $\alpha$ are most predictive of ADAPT-VQE performance.

\begin{figure}[t]
\includegraphics[width=0.48\textwidth,clip]{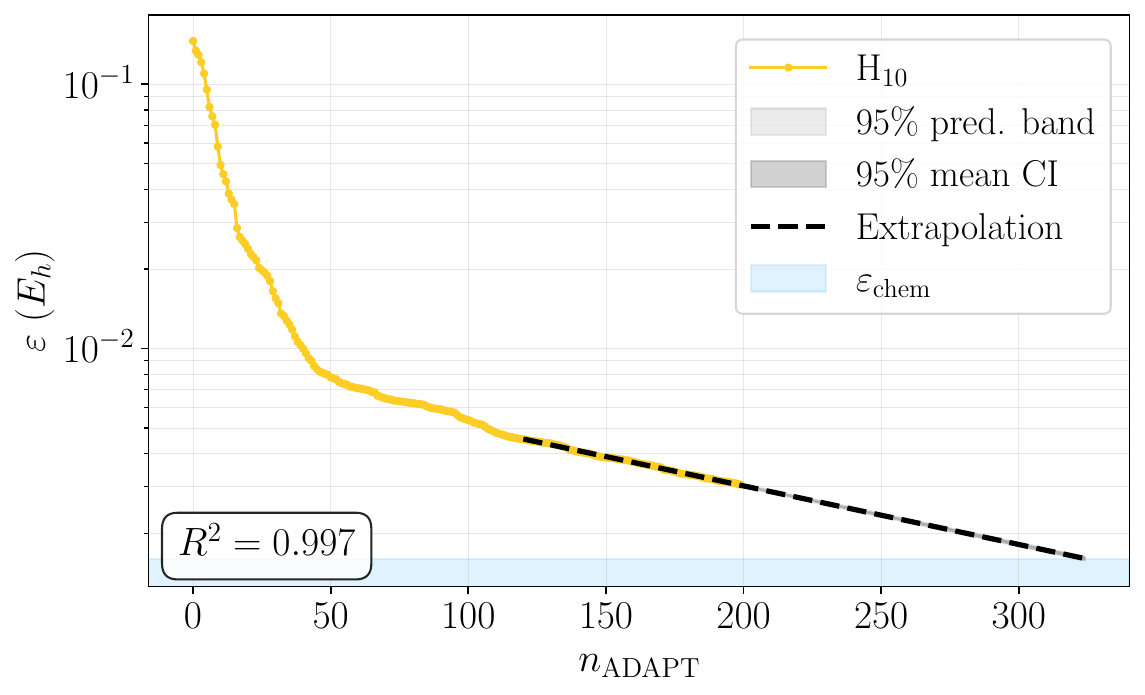}
\caption{\label{fig:h10_extrapolation} Extrapolation of the number of ADAPT-VQE iterations required to reach chemical accuracy for $\mathrm{H}_{10}$. A linear fit in the semi-logarithmic plane (dashed line) is performed in the range $120 \leq n_{\mathrm{ADAPT}} \leq 200$ and extrapolated to $\varepsilon_{\mathrm{chem}}$. Shaded regions indicate the 95\,\% prediction band (light) and the 95\,\% confidence interval of the mean (dark). Both are narrow enough to be barely visible, reflecting the high fit quality ($R^2 = 0.997$).}
\end{figure}
\begin{itemize} \item \textbf{Hartley entropy ($\alpha \to 0$).}
In this limit, \begin{equation} h_0 = \lim_{\alpha \to 0} h_\alpha = \ln N_{\mathrm{nz}}, \end{equation} where $N_{\mathrm{nz}}$ is the number of configurations with non-zero probability. This measure counts configurations without regard to their relative weights.

\item \textbf{Shannon entropy ($\alpha \to 1$).}  
The limit $\alpha \to 1$ yields the Shannon entropy,
\begin{equation}
    h_1 = \lim_{\alpha \to 1} h_\alpha = -\sum_i p_i \ln p_i,
\end{equation}
which weights each configuration by its probability. This measure captures the overall delocalization of the wave function across configurations.

\item \textbf{Collision entropy ($\alpha = 2$).}  
At $\alpha = 2$,
\begin{equation}
    h_2 = -\ln\!\left( \sum_i p_i^2 \right),
\end{equation}
which is related to the inverse participation ratio $\mathrm{IPR} = (\sum_i p_i^2)^{-1}$, a standard measure of wave function delocalization in quantum chemistry~\cite{MR.diagnostics.2015}. This choice emphasizes dominant configurations while suppressing the influence of small-probability contributions.

\item \textbf{Min-entropy ($\alpha \to \infty$).}  
In this limit,
\begin{equation}
    h_\infty = \lim_{\alpha \to \infty} h_\alpha = -\ln\!\left( \max_i p_i \right),
\end{equation}
which depends only on the largest CI coefficient. A value $h_\infty \approx 0$ indicates single-reference character (one dominant determinant), while larger values signal multi-reference character. This limiting case is conceptually related to the $T_1$ and $D_1$ diagnostics used in coupled-cluster theory, which also probe deviations from single-reference behavior~\cite{MR.diagnostics.2015}.

\end{itemize}

For ADAPT-VQE targeting chemical accuracy, the optimal order $\alpha \approx 0.25$ (Section~\ref{sec:quantifying_problem_complexity}) lies between these extremes: small enough to account for the tail of the distribution, yet large enough to avoid overweighting negligible configurations.
\begin{figure}[t]
\includegraphics[width=0.48\textwidth,clip]{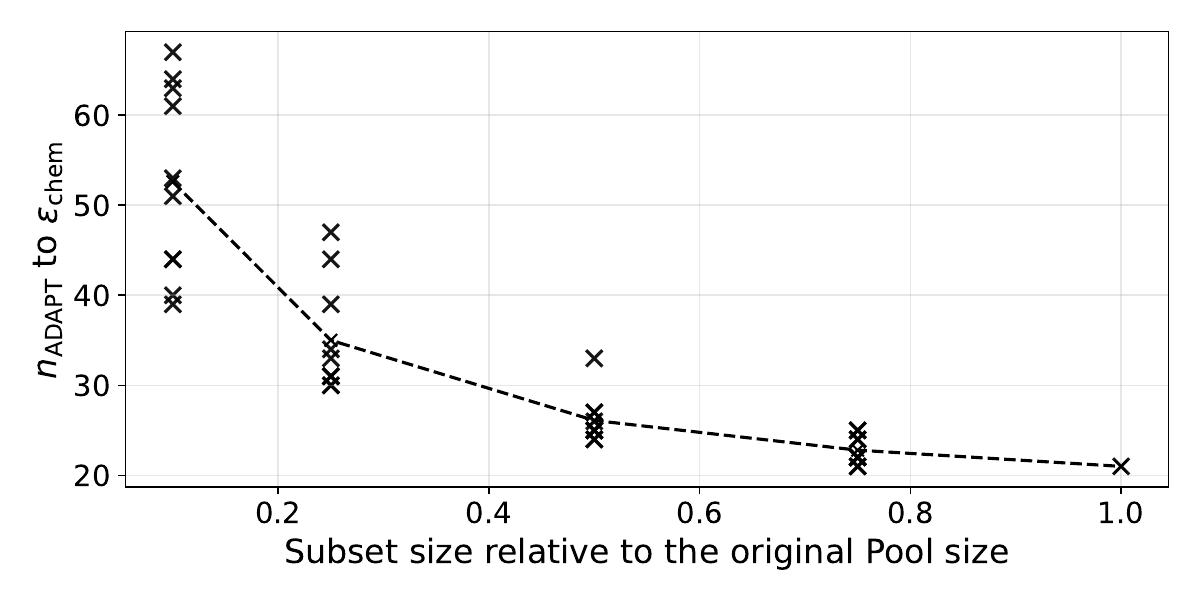}
\caption{\label{fig:mcp_h6} Number of ADAPT iterations $n_{\mathrm{ADAPT}}$ required to reach chemical accuracy as a function of operator-pool size (relative to the full pool from QEB-ADAPT-VQE) for H$_6$ at equilibrium geometry (1.0~\AA). For each pool-size ratio, ten random subsets were sampled; markers show individual runs and error bars indicate the standard deviation.}
\end{figure}

\section{\label{sec:extrapolation_h_10}Extrapolation of ADAPT Iterations for H$_{10}$}

As shown in Figure~\ref{fig:Hn_chain_equi}(a), H$_{10}$ does not reach chemical accuracy within 200 ADAPT iterations. However, for hydrogen chains larger than H$_8$, the convergence curves enter a linear regime in the semi-logarithmic plot near $\varepsilon_{\mathrm{chem}}$, enabling extrapolation.

We fit the H$_{10}$ data in the range $120 \leq n_{ADAPT} \leq 200$ to the functional form
\begin{equation}
    \log_{10}(\varepsilon) = a \cdot n_{\mathrm{ADAPT}} + b,
\end{equation}
obtaining $a = -0.002218 \pm 0.000028$ and $b = -2.0759 \pm 0.0044$ (95\% confidence intervals). Solving for $n_{\mathrm{ADAPT}}$ at $\varepsilon = \varepsilon_{\mathrm{chem}}$ yields $n_{\mathrm{ADAPT}} = 324.5 \pm 2.1$ iterations. Figure~\ref{fig:h10_extrapolation} shows this extrapolation; the fit residuals in the interpolation regime are small ($R^2 = 0.997$), supporting the reliability of the extrapolated value. This estimate is used in Figure~\ref{fig:Hn_chain_equi}(b) to establish the exponential scaling relationship.

\section{\label{sec:operator-pool_size}Operator-Pool Size and Spin-Symmetry Breaking}
Throughout this study, we observed that ADAPT-VQE transiently breaks spin symmetry during optimization: $\langle \hat{S}^2 \rangle$ deviates significantly from the expected eigenvalue in early iterations before recovering as convergence proceeds [Figure~\ref{fig:all_other_molecules}(c); see also Appendix~\ref{sec:h_additional_plots} for hydrogen chains]. This section examines whether this symmetry breaking is incidental or functionally important for optimization.

\begin{figure}[t]
\includegraphics[width=0.49\textwidth,clip]{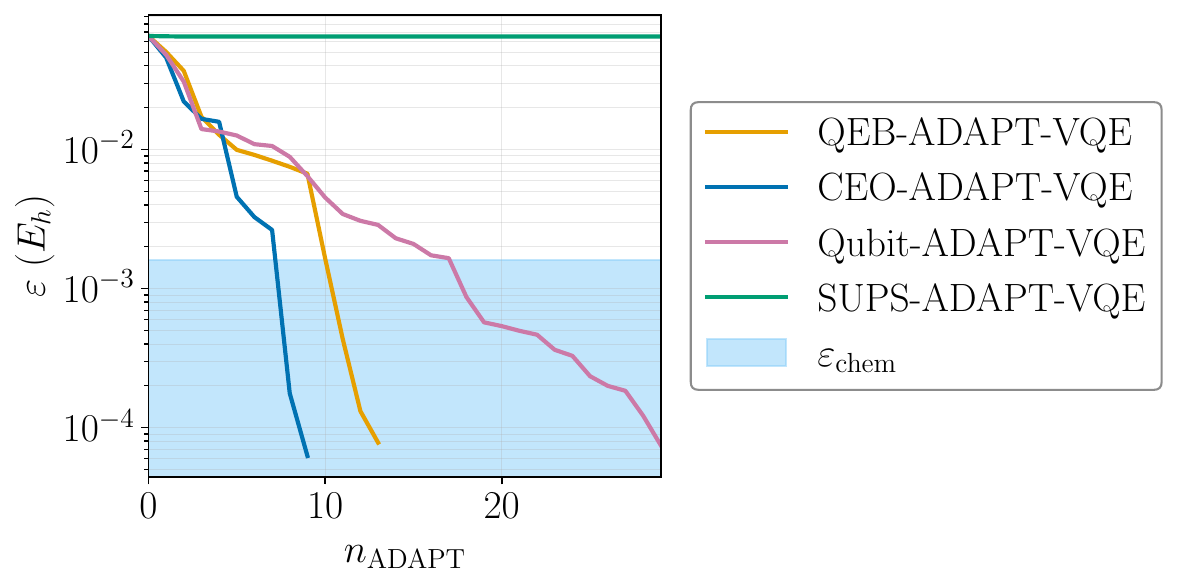}
\caption{\label{fig:sups_investigation} Convergence of ADAPT-VQE with the TETRIS extension for N$_2$ (1.2~\AA, (6,6) active space) using four operator pools: CEO, QEB, Qubit, and SUPS. The spin-preserving SUPS pool fails to converge, while all spin-non-preserving pools reach chemical accuracy within 20 iterations.}
\end{figure}
Recent work has demonstrated that operator pools can be reduced to sizes scaling linearly~\cite{mcp_paper} or quadratically~\cite{disco_vqe} with qubit count while retaining formal completeness—the ability to represent any state in Hilbert space given sufficient circuit depth. However, completeness does not guarantee that ADAPT-VQE can efficiently find a path to the target state. We investigated this trade-off by running QEB-ADAPT-VQE on H$_6$ (equilibrium geometry, 1.0~\AA) with randomly sampled subsets of the full operator pool.

Figure~\ref{fig:mcp_h6} shows the results: even at 10\% of the full pool size, chemical accuracy is achieved, but $n_{\mathrm{ADAPT}}$ approximately doubles compared to the full pool. The relationship is monotonic. Smaller pools consistently require more iterations. This indicates that while reduced pools remain formally complete, they constrain the optimization trajectory, forcing less direct paths to the ground state.

A more striking example arises from spin-preserving operator pools. The spin-adapted unitary pair-excitation singles (SUPS) pool~\cite{disco_vqe} scales as $\mathcal{O}(N^2)$ with qubit count $N$ and has been proven complete. Despite this, ADAPT-VQE fails to converge for molecules such as N$_2$ when using this pool~\cite{disco_vqe}. In contrast, the spin-non-preserving pools employed in this study (QEB, CEO, Qubit) converge within less than 20 iterations for the same system (Figure~\ref{fig:sups_investigation}).

This contrast suggests that transient spin-symmetry breaking plays a functional role in ADAPT-VQE optimization. The algorithm consistently selects symmetry-breaking operators in early iterations, enabling exploration of regions of the energy landscape inaccessible to spin-preserving trajectories. Symmetry is then restored as the optimization approaches the (spin-pure) ground state. Notably, even at chemical accuracy, residual $\langle \hat{S}^2 \rangle$ deviations can persist (up to $|\langle\hat{S}^2\rangle - S(S+1)| \approx 0.01$; see Figure~\ref{fig:all_other_molecules}(c)), which may introduce systematic errors if VQE-optimized states are used in workflows that assume spin purity, such as CASSCF orbital optimization.

These observations highlight a trade-off in operator-pool design: pools that preserve symmetries or minimize size may sacrifice optimization efficiency, while larger symmetry-breaking pools provide more direct convergence paths at the cost of increased gradient evaluation overhead.

\begin{figure}[tb]
\includegraphics[width=0.49\textwidth,clip]{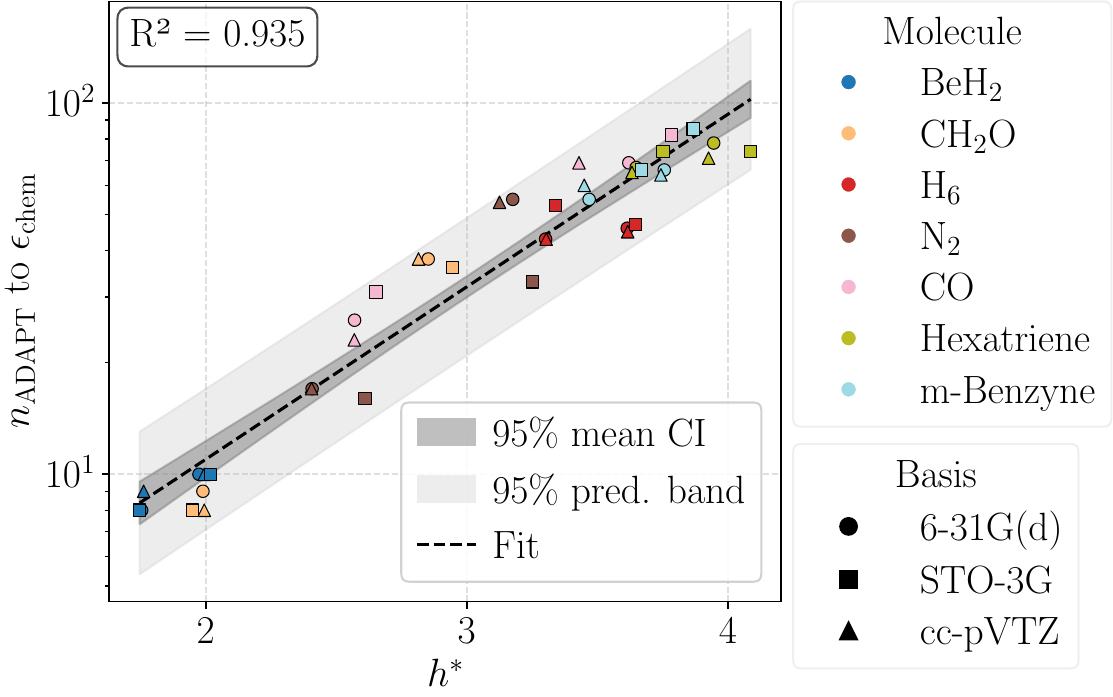}
\caption{\label{fig:renyi_6_orbs_fit} Correlation between the Rényi entropy $h^\ast$ and the number of ADAPT iterations $n_{\mathrm{ADAPT}}$ required to reach chemical accuracy $\varepsilon_{\mathrm{chem}}$. Here, $h^\ast \equiv h_{\alpha^\ast}$ with $\alpha^\ast \approx 0.25$ as defined in Section~\ref{sec:quantifying_problem_complexity}. Data include all six-orbital systems from Table~\ref{tab:molecule_set} in the listed basis sets, each in the singlet ground state and first excited triplet state, computed with QEB-ADAPT-VQE without the TETRIS extension. The linear fit (dashed line) achieves R$^2 = 0.94$.}
\end{figure}
\section{\label{sec:renyi_6_orbs_fit}Rényi Entropy Correlation for Six-Orbital Molecules}
Figure~\ref{fig:renyi_6_orbs_fit} shows a cross-section of the R$^2$ surface from Figure~\ref{fig:r2_surface_renyi} at $\varepsilon_{\mathrm{chem}}$. As defined in Section~\ref{sec:quantifying_problem_complexity}, $h^\ast$ denotes the Rényi entropy evaluated at the order $\alpha$ that maximizes R$^2$ for a given target error; at $\varepsilon_{\mathrm{chem}}$, this optimal order is $\alpha \approx 0.25$, yielding $h^\ast = h_{0.25}$.

\begin{figure*}[t]
\includegraphics[width=0.99\textwidth,clip]{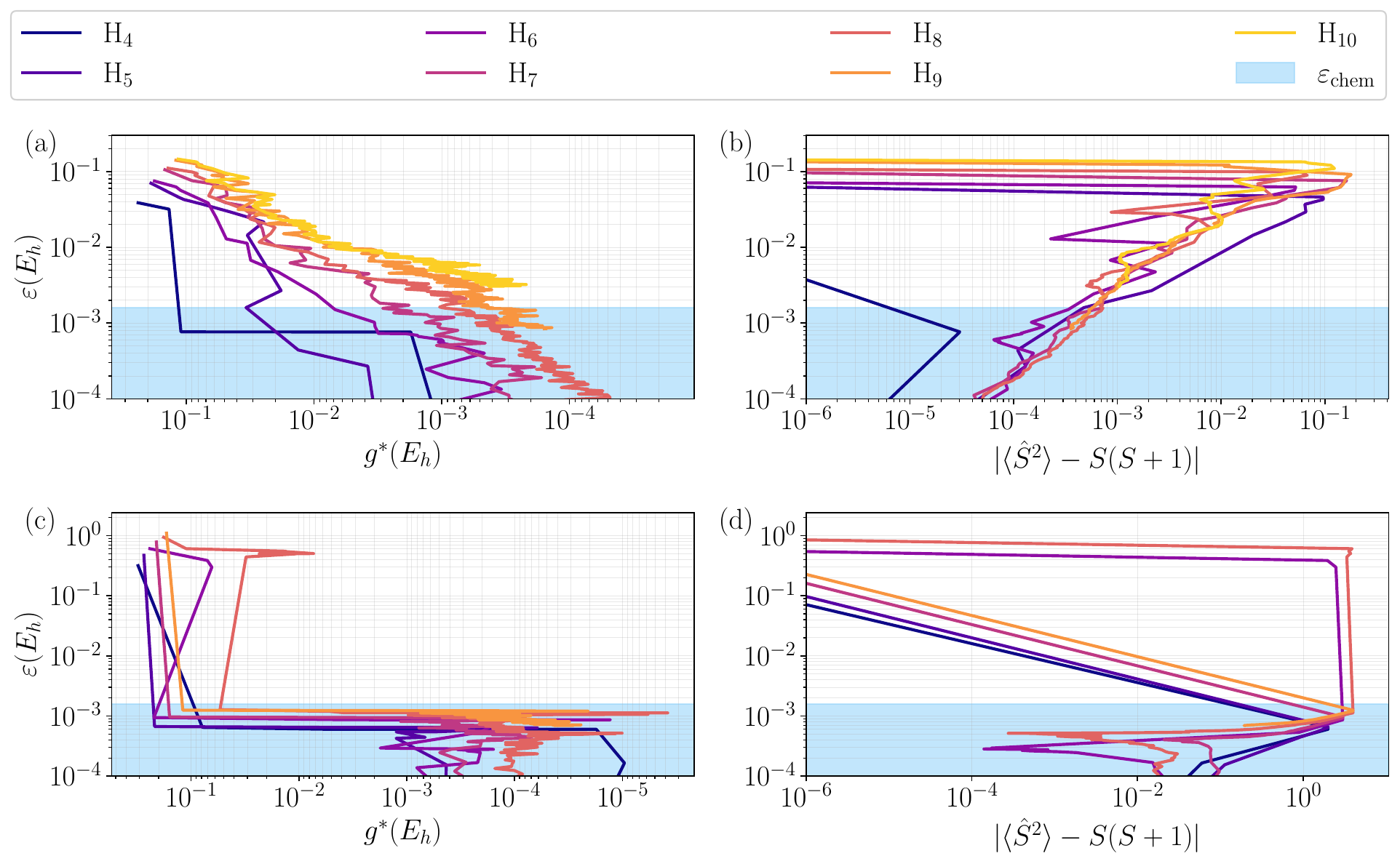}
\caption{\label{fig:Hn_chain_further} Convergence diagnostics for hydrogen chains H$_n$ ($n = 4$--$10$ for equilibrium; $n = 4$--$9$ for stretched). Top row: equilibrium geometry (1.0~\AA); bottom row: stretched geometry (3.0~\AA). (a, c) Largest operator gradient magnitude $g^\ast$ versus energy error $\varepsilon$. (b, d) Spin deviation $|\langle\hat{S}^2\rangle - S(S+1)|$ versus $\varepsilon$, where $S$ is the total spin quantum number (0 for singlet, 1/2 for doublet). In blue chemical accuracy $\varepsilon_{\mathrm{chem}}$ is indicated. All calculations use CEO-ADAPT-VQE with the TETRIS extension.}
\end{figure*}
Each point in Figure~\ref{fig:renyi_6_orbs_fit} represents a six-orbital molecule from Table~\ref{tab:molecule_set} in a specific basis set and spin state. Despite all systems having identical active space size (and therefore qubit count), the required number of ADAPT iterations $n_{\mathrm{ADAPT}}$ spans approximately one order of magnitude, from roughly 8 to 80 iterations. This variation underscores that active space size alone is an inadequate predictor of VQE complexity; the Rényi entropy captures the relevant differences in multi-reference character that determine convergence behavior.

\section{\label{sec:h_additional_plots}Additional Convergence Data for Hydrogen Chains}
Figure~\ref{fig:Hn_chain_further} shows the gradient magnitudes and spin expectation values during ADAPT-VQE optimization for hydrogen chains at equilibrium and stretched geometries, complementing the energy convergence data in Figure~\ref{fig:Hn_chain_equi} and Figure~\ref{fig:Hn_chain_stretched_iter}.

Figure~\ref{fig:Hn_chain_further}(a) shows the largest operator gradient $g^\ast$ during optimization for equilibrium-geometry hydrogen chains. While the gradients exhibit larger fluctuations than the energy convergence, they decrease toward zero as expected for a converging optimization. Notably, at a given energy error $\varepsilon$, larger systems exhibit smaller gradient magnitudes. We attribute this to the deeper circuits required to achieve the same $\varepsilon$ in larger systems. This observation suggests that ADAPT-VQE may encounter diminishing gradients for increasingly large systems in practice, a consequence of the exponentially growing circuit depth identified in Section~\ref{sec:h_chains_equilibrium}. However, the fluctuations and limited data points preclude quantitative scaling analysis.

Figure~\ref{fig:Hn_chain_further}(c) shows the corresponding gradient data for stretched geometries. Similar trends are observed, but the step-wise energy convergence produces larger discontinuities in the gradient magnitudes, further obscuring any systematic size dependence.

Figure~\ref{fig:Hn_chain_further}(b) shows the spin deviation $|\langle\hat{S}^2\rangle - S(S+1)|$ during optimization for equilibrium-geometry chains. Except for H$_4$, all optimizations initially select operators that produce significant spin contamination ($|\langle\hat{S}^2\rangle - S(S+1)| \approx 0.1$), which then diminishes as the energy converges. The absence of transient spin contamination for H$_4$ may reflect the simplicity of this system, where spin-symmetry breaking is unnecessary to find an efficient optimization trajectory.

Figure~\ref{fig:Hn_chain_further}(d) shows the spin deviation for stretched geometries. The initial contamination is substantially larger ($|\langle\hat{S}^2\rangle - S(S+1)| > 1$) and, crucially, persists even after chemical accuracy is reached. This indicates that the algorithm has converged to a state close in energy to the target but with incorrect spin character, consistent with the step-wise convergence behavior discussed in Section~\ref{sec:stretched_h_chains}. For stretched geometries, the spin deviation provides valuable diagnostic information beyond the energy error: significant residual spin contamination signals that the true target state has not been properly recovered, even when $\varepsilon < \varepsilon_{\mathrm{chem}}$.

\begin{figure}[tb]
\includegraphics[width=0.49\textwidth,clip]{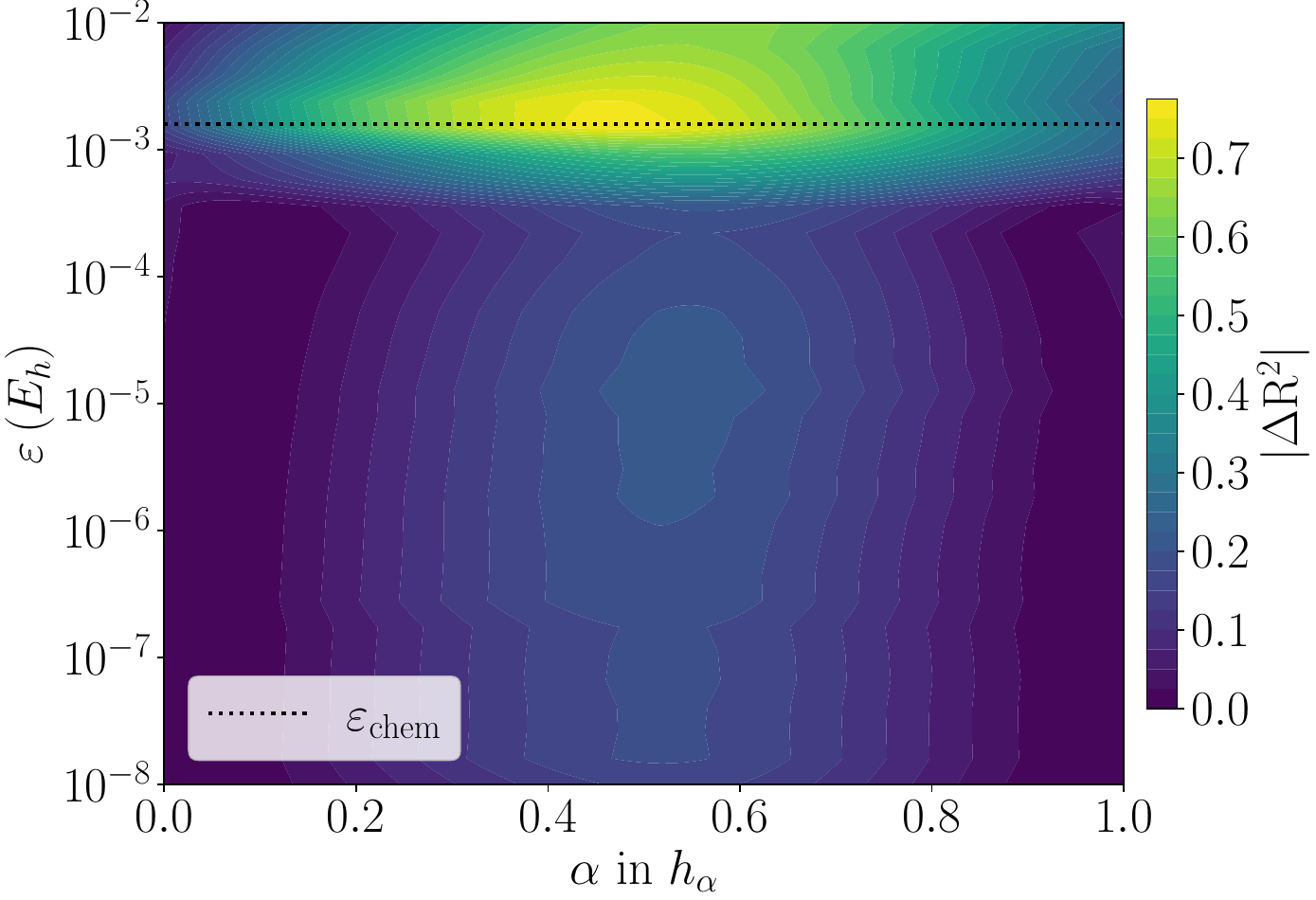}
\caption{\label{fig:renyi_stretched} Absolute difference in R$^2$ between linear fits of $\log(n_{\mathrm{ADAPT}})$ versus $h_\alpha$ with and without inclusion of H$_6$ at stretched geometry (3.0~\AA). The baseline fit uses all six-orbital molecules from Table~\ref{tab:molecule_set} at equilibrium geometry (three basis sets, singlet and triplet states each; CEO-ADAPT-VQE with TETRIS). Large differences at high $\varepsilon$ indicate that stretched-geometry convergence behavior deviates from equilibrium trends; the difference vanishes for $\varepsilon \lesssim 5 \times 10^{-4}~E_h$.}
\end{figure}
\section{\label{sec:renyi_stretched}Rényi Entropy Prediction for Stretched Geometries}
Section~\ref{sec:stretched_h_chains} established that stretched hydrogen chains exhibit qualitatively different convergence behavior (step-wise and non-monotonic) compared to equilibrium geometries. Here we examine whether the Rényi entropy remains predictive for stretched geometries by quantifying how including stretched-geometry data affects the quality of the correlation.

Figure~\ref{fig:renyi_stretched} shows the absolute difference in R$^2$ between fits that include versus exclude H$_6$ at stretched geometry (3.0~\AA). At large $\varepsilon$, the difference is substantial (up to 0.15), indicating that stretched-geometry data points deviate significantly from the trend established by equilibrium molecules. This deviation reflects the step-wise convergence behavior: at modest iteration counts, the energy error depends sensitively on whether the optimization has crossed a plateau, making $n_{\mathrm{ADAPT}}$ poorly predicted by smooth correlations.

As $\varepsilon$ decreases, the R$^2$ difference approaches zero. Below $\varepsilon \approx 5 \times 10^{-4}~E_h$, the stretched-geometry data integrate seamlessly with the equilibrium trend. This threshold corresponds to the regime where stretched H$_6$ has traversed its final plateau and entered smooth convergence toward the target state (Figure~\ref{fig:Hn_chain_stretched_iter}). In this asymptotic regime, the step-wise features average out and the Rényi entropy recovers its predictive power.

These observations justify our focus on equilibrium geometries for establishing scaling relationships (Section~\ref{sec:quantifying_problem_complexity}): the predictable, monotonic convergence at equilibrium provides a reliable foundation for extrapolation, whereas stretched geometries introduce irregularities that complicate systematic analysis at practically relevant accuracy thresholds. Nevertheless, it appears that the Rényi entropy remains predictive for stretched geometries when sufficiently stringent error thresholds are targeted, indicating that the correlation reflects a fundamental relationship rather than one limited to equilibrium configurations.

\begin{figure}[t]
\includegraphics[width=0.49\textwidth,clip]{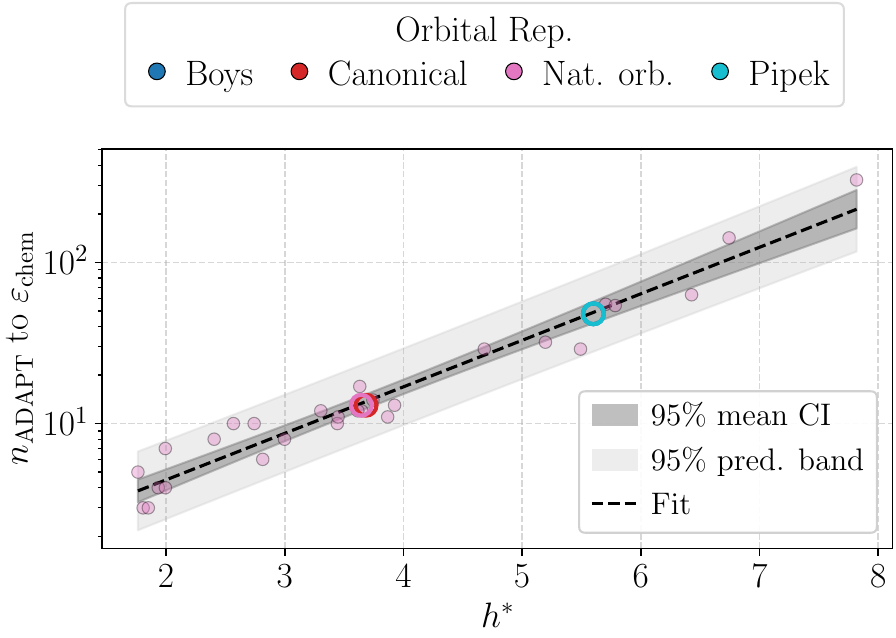}
\caption{\label{fig:orb_rep_comparison} Correlation between the Rényi entropy $h^\ast$ and the number of ADAPT iterations $n_{\mathrm{ADAPT}}$ required to reach chemical accuracy for H$_6$ (1.0~\AA, singlet) in different orbital representations: Boys, canonical, natural, and Pipek orbitals (rings). Boys and Pipek orbitals result in very similar $h^\ast$ and $n_{\mathrm{ADAPT}}$ so that the difference can not be resolved. Here, $h^\ast \equiv h_{\alpha^\ast}$ with $\alpha^\ast \approx 0.25$ as defined in Section~\ref{sec:quantifying_problem_complexity}. Calculations performed with CEO-ADAPT-VQE with the TETRIS extension. For comparison the same data (circles) and fit as in Figure~\ref{fig:paper_overview_sketch}(b) are shown.}
\end{figure}
\section{\label{sec:renyi_other_orb_reps}Rényi Entropy in Different Orbital Representations}
The Rényi entropy and ADAPT-VQE performance both depend on the choice of orbital representation, since this choice affects the sparsity of the CI expansion and the structure of the optimization landscape. Figure~\ref{fig:orb_rep_comparison} compares different orbital representations for H$_6$ at equilibrium geometry.

Natural orbitals yield both the lowest Rényi entropy and the fewest ADAPT iterations, while Pipek orbitals and Boys orbitals require approximately four times as many iterations. The data points for all four representations agree closely with the linear fit from Figure~\ref{fig:paper_overview_sketch}(b) on all molecules, indicating that our calibrated relationship between $h^\ast = h_{0.25}$ and $n_\mathrm{ADAPT}$ generalizes across orbital choices.

These observations have two practical implications. First, natural orbitals should be preferred for VQE calculations, as they minimize both the multi-reference character and the required circuit depth. Second, our scaling predictions based on natural orbitals (Sections~\ref{sec:quantifying_problem_complexity} and~\ref{sec:extrapolation}) represent favorable estimates; calculations in other orbital representations would require more iterations but would still follow the same exponential scaling relationship.

\begin{figure}[t]
\includegraphics[width=0.49\textwidth,clip]{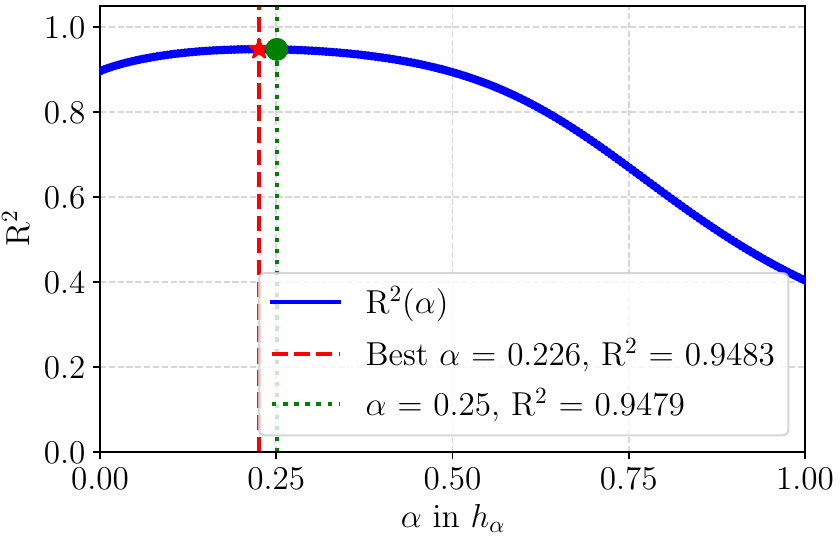}
\caption{\label{fig:best_alpha_chem_acc_all_molecules} Coefficient of determination (R$^2$) for the linear fit between $\log(n_{\mathrm{ADAPT}})$ and the Rényi entropy $h_\alpha$ as a function of the order $\alpha$, evaluated at chemical accuracy $\varepsilon_{\mathrm{chem}}$. Data include all molecules from the MolVQE-21 benchmark set (Table~\ref{tab:molecule_set}) using CEO-ADAPT-VQE with the TETRIS extension. The green vertical dashed line indicates $\alpha^\ast \approx 0.25$, the optimal value determined from six-orbital molecules (Section~\ref{sec:quantifying_problem_complexity}), while the red vertical dashed line indicates the $\alpha$ recovering the highest R$^2$ score on the full MolVQE-21 benchmark set.}
\end{figure}
\section{\label{sec:alpha_scan_all_molecules}Optimal Rényi Order Across the Full Benchmark Set}
Section~\ref{sec:quantifying_problem_complexity} established $\alpha = 0.25$ as optimal for six-orbital molecules at chemical accuracy. Figure~\ref{fig:best_alpha_chem_acc_all_molecules} tests whether this choice generalizes to the full MolVQE-21 benchmark set by scanning $\alpha$ and computing the R$^2$ score for the correlation between $h_\alpha$ and $\log(n_{\mathrm{ADAPT}})$.

The R$^2$ curve exhibits a broad plateau around its maximum, with $\alpha = 0.25$ achieving R$^2 = 0.948$, only $\Delta\mathrm{R}^2 = 0.0004$ below the global maximum. This confirms that the optimal order calibrated on six-orbital molecules transfers to the full benchmark set without retuning. The stability of the plateau further indicates that moderate variations in $\alpha$ (e.g., $\alpha \in [0.2, 0.3]$) yield nearly equivalent predictive power, obviating the need for precise optimization of this parameter for interpolation.

\begin{figure*}[t]
\includegraphics[width=0.99\textwidth,clip]{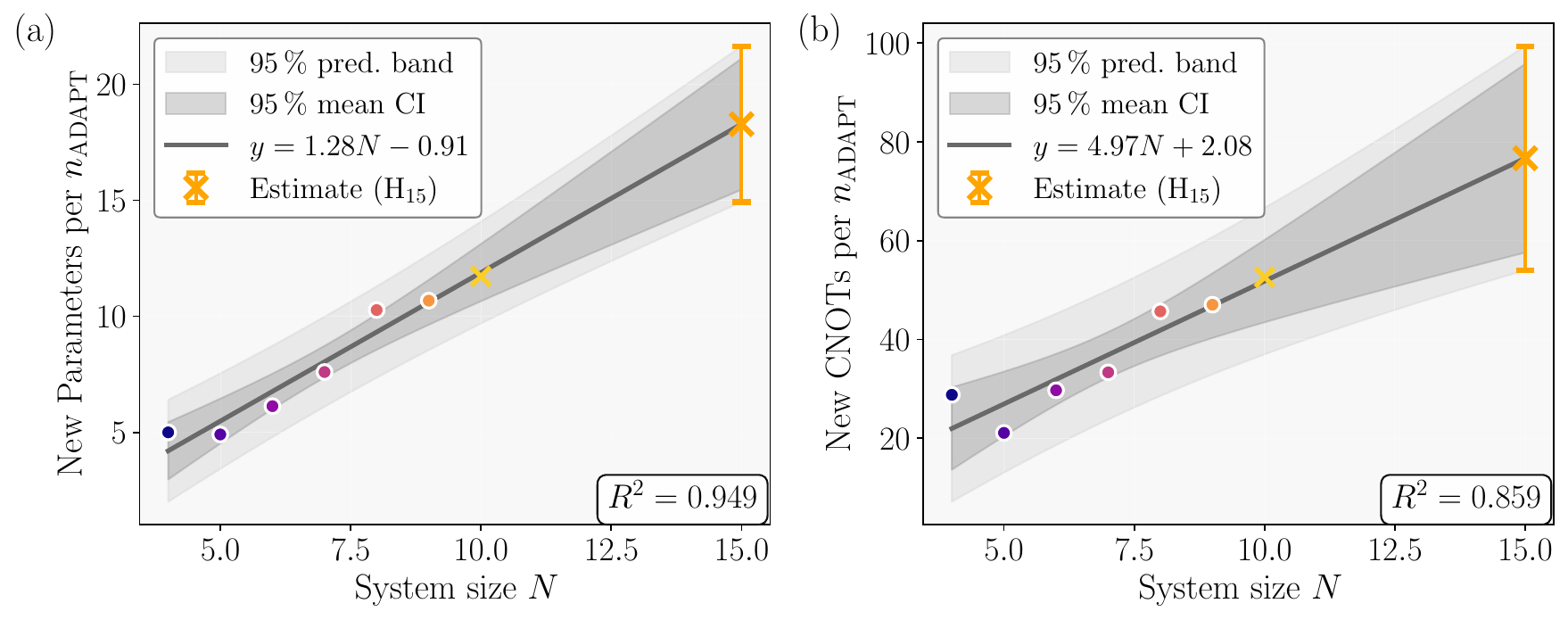}
\caption{\label{fig:combined_extrapolation}%
Extrapolation of per-iteration ADAPT-VQE circuit resources to $\mathrm{H}_{15}$ based on hydrogen chains $\mathrm{H}_4$--$\mathrm{H}_{10}$. (a)~Mean number of new variational parameters per ADAPT-VQE iteration. (b)~Mean number of new CNOT gates per iteration. In both panels, shaded regions indicate the 95\,\% prediction band (light) and the 95\,\% confidence interval of the mean (dark), and orange crosses mark the extrapolated estimates for $\mathrm{H}_{15}$.}
\end{figure*}
\section{\label{sec:cnot_parameter_estimation}Derivation of CNOT Gate and Parameter Estimates}
To estimate the total number of parameters and CNOT gates required for H$_{15}$, we decompose each total as $n_{\mathrm{total}} = n_{\mathrm{ADAPT}} \times r$, where $r$ denotes the mean number of new parameters (or CNOTs) added per ADAPT iteration. We estimate $r(N{=}15)$ via linear regression on the per-iteration rates observed for H$_4$ through H$_{10}$ (Figure~\ref{fig:combined_extrapolation}).

The per-iteration parameter and CNOT counts scale approximately linearly with system size [Figure~\ref{fig:combined_extrapolation}(a) and (b)]. This scaling is a direct consequence of the TETRIS extension: standard ADAPT-VQE appends a single operator per iteration (yielding roughly constant per-iteration cost), whereas TETRIS appends multiple operators acting on disjoint qubit subsets simultaneously. The number of such non-overlapping operators that can be selected in parallel grows linearly with the number of qubits, producing the observed linear scaling of $r$ with system size.

Notable fluctuations around the linear trend are visible. When the system size increases from an odd to an even number of atoms, many new double-excitation operators become accessible and contribute significantly to each iteration, whereas a transition from even to odd primarily introduces single excitations, which carry comparatively smaller circuit overhead. This behavior reflects the interplay between TETRIS parallelization and operator availability. Additional fluctuations, particularly for small systems, can be attributed to the distinction between single- and multi-parameter operators in the CEO-ADAPT-VQE ansatz. 

Despite these fluctuations, the linear trend provides a reasonable fit (R$^2 = 0.949$ for parameters; R$^2 = 0.859$ for CNOTs), and multiplying the extrapolated per-iteration rates by the predicted iteration count (Section~\ref{sec:extrapolation}) yields our estimates of the total parameter and CNOT budgets for H$_{15}$.

\section{\label{sec:renyi_scaling}Scaling Analysis via Rényi Entropy}

The linear relationship between $h_\alpha$ and $\log(n_\mathrm{ADAPT})$ observed in Figure~\ref{fig:paper_overview_sketch}(b) implies
\begin{equation}
    n_\mathrm{ADAPT}(\varepsilon) \propto \exp(c \cdot h_\alpha) = \left( \sum_i p_i^\alpha \right)^{c/(1-\alpha)},
\end{equation}
where $c > 0$ is the slope of the linear fit. This relationship connects the iteration count to the structure of the CI coefficient distribution through the sum $\sum_i p_i^\alpha$.

For approximate state recovery (large $\alpha$), only the largest probabilities contribute significantly to this sum:
\begin{equation}
    0 < \sum_i p_i^\alpha \leq 1 \quad \text{for } \alpha \gg 1,
\end{equation}
yielding bounded iteration requirements independent of system size.

For high-fidelity reconstruction ($\alpha \to 0$), the sum approaches the number of configurations with non-zero probability:
\begin{equation}
    \lim_{\alpha \to 0} \sum_i p_i^\alpha = N_{\mathrm{nz}},
\end{equation}
where $N_{\mathrm{nz}}$ is the number of determinants with non-zero CI coefficients. This quantity is upper bounded by the total number of determinants in the active space:
\begin{equation}
    N_{\mathrm{nz}} \leq N_{\mathrm{det}} = \binom{n_{\mathrm{orb}}}{n_\alpha} \binom{n_{\mathrm{orb}}}{n_\beta} = \mathcal{O}\!\left(4^{n_{\mathrm{orb}}}\right),
\end{equation}
where $n_\alpha$ and $n_\beta$ are the numbers of alpha and beta electrons, respectively, $n_{\mathrm{orb}}$ is the number of active orbitals, and the final scaling assumes half-filling ($n_\alpha \approx n_\beta \approx n_{\mathrm{orb}}/2$).

This analysis indicates that for high-fidelity state reconstruction, the number of ADAPT iterations scales with the number of significantly populated configurations, which in the worst case grows exponentially with the number of active orbitals. This theoretical expectation is consistent with the empirical exponential scaling observed for hydrogen chains in Section~\ref{sec:h_chains_equilibrium}.

\section{\label{sec:molecular_geometry_optimization}Molecular Geometry Optimization}
For molecules where literature geometries were unavailable or unsuitable for our active space choices, we performed geometry optimizations at the CASSCF level using PySCF's interface to the geometric optimizer PyBerny~\cite{pyberny}. Optimizations were performed with the same active space and basis set used in subsequent VQE calculations, ensuring consistency between the nuclear configuration and the electronic structure treatment. For Hexatriene the starting geometry was taken from~\cite{hexatriene_ref} and for M-Benzyne was taken from~\cite{m_benzyne_ref}.
\end{document}